\documentclass[aps,prb,twocolumn,floatfix,superscriptaddress,longbibliography]{revtex4-1}
\usepackage{graphicx,txfonts,bm}
\usepackage[colorlinks,citecolor=magenta,linkcolor=blue]{hyperref}
\usepackage{hyperref}
\usepackage{natbib}

\begin{document}

\title{Orbital-selective correlations and angular momentum coupling in heavy actinides Am, Cm, Bk, and Cf under pressure: A many-body perspective}
\author{Haiyan Lu}
\email{hyluphys@163.com}
\affiliation{Institute of Materials, China Academy of Engineering Physics, Huafengxincun No.~9, Mianyang 621907, Sichuan, China}

\begin{abstract}
We systematically investigate the electronic structures of americium (Am), curium (Cm), berkelium (Bk), and californium (Cf) in both the ambient-pressure double hexagonal close-packed (dhcp) and high-pressure face-centered cubic (fcc) phases, using density functional theory combined with embedded dynamical mean-field approach. Our results reveal that Am exhibits moderate correlation strength and localized 5$f$ states dominated by $jj$ angular momentum coupling scheme. In Cm and Bk, strong electron correlations drive the system into a localized regime, characterized by Hubbard band formation, large effective electron masses, and non-Fermi liquid behavior. Their magnetic ground states are governed by exchange interactions within an intermediate coupling scheme that shifts toward LS coupling. Remarkably, Cf reenters a $jj$ coupling regime while exhibiting the strongest orbital-selective correlations among the series. Atomic eigenstate probabilities show moderate configurational mixing in Am, whereas Cm, Bk, and Cf maintain nearly fixed trivalent configurations, indicating localized 5$f$ states. Compared with the dhcp phase, the fcc structure generally enhances correlation effects, as evidenced by wider Hubbard bandgaps and increased valence state fluctuation in Am. Analyses of kinetic energy, potential energy, spin susceptibility, and charge susceptibility further corroborate the progressive localization of 5$f$ electrons and the emergence of orbital-selective correlations from Am to Cf. This work establishes a unified picture of 5$f$ electron evolution across the Am-Cf series, elucidating the interplay between spin–orbit coupling, electron correlation, and crystal structure in heavy actinides and offering insights into their behavior under high pressure.
\end{abstract}

\maketitle
\section{Introduction\label{sec:introduction}}

Actinides have attracted significant interest in both physics and chemistry due to the unique physical properties arising from their partially filled 5$f$ subshells. In lighter actinides, such as uranium (U) and neptunium (Np), 5$f$ electrons exhibit itinerancy, while plutonium (Pu) lies at the boundary between itinerant and localized 5$f$ states. In stark contrast, for transplutonium elements specifically americium (Am), curium (Cm), berkelium (Bk), and californium (Cf), the 5$f$ electrons become increasingly localized with rising atomic number. This trend leads to predominantly localized 5$f$ states in these heavier elements, resulting in electronic structures and physical properties that are distinctly different from those of their lighter counterparts~\cite{RevModPhys.81.235,RevModPhys.98.015004}. In transplutonium elements, the partially filled 5$f$ shells give rise to strong electron correlations, which intertwine with the inherent strong spin-orbit coupling (SOC) characteristic of nuclei and the crystal field splitting imposed by complex crystal structures. Together, these factors produce exceptionally intricate electronic structures and lead to abundant novel quantum phenomena~\cite{Science1978AmSC,PhysRevLett.94.097002,Science.309.110,PhysRevB.87.214111}.

Am, Cm, Bk, and Cf constitute the first four transplutonium elements, spanning a gradual increase in 5$f$ electron count from $5f^6$ to $5f^{10}$ and capturing the crossover from non-magnetic local-moment behavior to valence fluctuations.
Am represents the first transplutonium actinide in which the 5$f$ electrons form a closed relativistic subshell. Under strong spin–orbit coupling, the 5$f$ orbitals split into $j$ = 5/2 and $j$ = 7/2 manifolds. The six 5$f$ electrons predominantly occupy the $j$ = 5/2 subshell, resulting in a non-magnetic, atomic-like 5$f^{6}$ configuration. This produces a $^{7}F_0$ singlet ground state within the $jj$ angular momentum coupling scheme, characterized by total angular momentum $J=0$~\cite{PhysRevLett.94.097002,PhysRevB.76.073105}. Under high pressure, Am exhibits four allotropes under pressures up to 100~GPa~\cite{heathman2000,PhysRevB.63.214101,Lindbaum2003}. Especially, superconductivity emerges near 0.8~K at ambient pressure, reaches a maximum $T_c$ of 2.3~K around 10~GPa, and then declines upon further compression~\cite{Science1978AmSC,LINK1994148,GRIVEAU200784}, making it the only experimentally confirmed transplutonium superconductor to date. Despite extensive studies,a central unresolved question concerns whether and under what pressure the 5$f$ electrons become itinerant upon compression~\cite{PhysRevB.82.201103,heathman2000,PhysRevB.63.214101,EPL82.57007,JPCM17.257,PhysRevB.72.024109}.

Cm exhibits five allotropes and undergoes four successive phase transitions up to 100~GPa~\cite{Science.309.110,Cmhighpressure2007}. These transitions involve significant volume collapses of approximately 4.5\% for the Cm (II-III) transition and 11.7\% for Cm (IV-V) transition, reflecting the delocalization of 5$f$ electrons under compression. Typically, Cm deviates toward the intermediate coupling regime with a strong bias toward LS coupling, as evidenced by electron energy-loss spectroscopy, x-ray absorption spectroscopy and x-ray magnetic circular dichroism measurements~\cite{PhysRevLett.98.236402,PhysRevB.76.073105,Muller2021Cf}. The seven 5$f$ electrons form a half-filled shell stabilized by exchange interaction, resulting in a large spin moment and a magnetic ground state~\cite{PhysRevLett.98.236402,Science.309.110}. However, resonant x-ray experiments have failed to detect any diffraction peaks, leaving the antiferromagnetic ground state of Cm an open question~\cite{PhysRevB.99.224419}.

The initial dhcp structure of Bk first transforms to the fcc structure at 8~GPa, which above 22~GPa converts to an orthorhombic $\alpha$-uranium-type structure, followed by a sudden volume decrease of about 12\%. This collapse indicates the onset of 5$f$ electrons delocalization and their active participation in bonding under high pressure~\cite{HAIRE1984119,Benedict1984}. Furthermore, Bk falls into the intermediate coupling regime with a shift towards LS coupling, as revealed by electron energy-loss spectroscopy measurements~\cite{Muller2021Cf}. In this regime, exchange interactions dominate over the comparatively weak spin–orbit coupling, leading to a magnetic ground state characterized by a large spin moment arising from the parallel alignment of spins according to Hund's rule~\cite{RevModPhys.81.235}.

Cf also possesses a rich phase diagram, exhibiting three progressive high-pressure phases up to 100~GPa~\cite{PhysRevB.87.214111,Inorg.Nucl.Chem.36.1295}. Notably, Cf exhibit distinct valence state fluctuations, including divalent, trivalent, and intermediate valence states~\cite{PhysRevB.99.045109}. Under applied pressure, the decreasing interatomic distances enhance hybridization between the initially localized 5$f$ electrons and conduction electrons, thereby promoting a gradual transition toward itinerancy. This process is often coupled with significant structural phase transitions and volume collapses, stabilizing low-symmetry structures such as monoclinic and orthorhombic phases at high pressures~\cite{Science.309.110,PhysRevLett.85.2961,PhysRevB.87.214111}. More broadly, theoretical predictions and experimental indications suggest that transplutonium systems may harbor a wealth of exotic quantum states, including heavy-fermion behavior, valence state fluctuations, quantum criticality, and topological phases~\cite{nature4202002,shim:2007,PhysRevLett.111.176404,bauerreview}.

Experimental investigations of transplutonium elements face formidable challenges due to their intense radioactivity, severe chemical toxicity, high reactivity, and extreme difficulty of sample preparation. These materials can only be synthesized in minute quantities and at prohibitively high costs~\cite{LAReview}. Consequently, our understanding of the condensed matter physics of transplutonium elements remains severely limited.
To date, experimental efforts have made progress on high-pressure crystal structures by using synchrotron x-ray diffraction~\cite{Science.309.110,PhysRevLett.85.2961,PhysRevB.87.214111,RevModPhys.98.015004}. In addition, a variety of spectroscopic techniques including x-ray absorption spectroscopy (XAS), inelastic x-ray scattering, x-ray magnetic circular dichroism (XMCD), electron energy-loss spectroscopy (EELS), electrical resistivity, infrared spectroscopy, photoluminescence, Raman spectroscopies, and photoemission spectroscopy have been applied to probe charge distribution, density of states, crystal field splitting, magnetic properties, and atomic configuration distributions~\cite{PhysRevB.82.201103,PhysRevB.99.224419,PhysRevB.77.113109,PhysRevB.76.073105,PhysRevLett.98.236402,Muller2021Cf}.
These investigations have yielded several key findings. Temperature-driven phase transitions have been identified in Am and Cm, and high-pressure structural phases have been characterized for Am, Cm, Bk, and Cf~\cite{heathman2000,PhysRevB.63.214101,Science.309.110,Cmhighpressure2007,HAIRE1984119,PhysRevB.87.214111}. Superconductivity has been observed in Am~\cite{Science1978AmSC,LINK1994148,GRIVEAU200784}, while magnetic properties have been explored in Cm and Bk~\cite{PhysRevLett.98.236402,RevModPhys.81.235}. Furthermore, pressure-induced valence state fluctuations have been reported in Cf~\cite{RevModPhys.81.235,PhysRevB.99.045109,PhysRevB.87.214111}. In the realm of chemistry, breakthroughs have elucidated the complex bonding behavior of elements such as Cf~\cite{Goodwin2021}.
Despite these advances, direct experimental data remain scarce for many fundamental physical properties of transplutonium elements, including electronic band structures, Fermi surfaces, mechanical behavior, thermodynamic properties, lattice dynamics, optical responses, and magnetic ordering~\cite{RevModPhys.81.235,PhysRevLett.94.097002,PhysRevB.99.224419,RevModPhys.95.015001}.

The strongly correlated 5$f$ electrons in transplutonium elements challenge conventional density functional theory (DFT) and its DFT+$U$ extensions within a single-particle picture, which fail to reproduce the experimental photoemission spectrum of Am-I~\cite{PhysRevB.84.075138,PhysRevB.73.104415,J.AlloysCompd.444.42,SolidStateCommun.150.938,SolidStateCommun.164.22} and frequently predict magnetic ground states for Am-I and Am-II, contradicting the observed non-magnetic behavior~\cite{JPCM14.3575,JPCM17.257,PhysRevB.72.024109,PhysRevB.45.3198}. Hybrid density functionals~\cite{Chem.Phys.Lett.482.223} improve the agreement with photoemission data~\cite{PhysRevB.72.115122}, yet they remain insufficient for fully capturing the many-body correlations inherent to 5$f$ electrons. For Cm, the lowest energy of the antiferromagnetic state using the full potential linear muffin-tin orbital (FPLMTO) method shows that spin polarization of the 5$f$ electrons stabilizes the high-pressure phase Cm-III~\cite{Science.309.110}.

A more sophisticated framework, DFT combined with dynamical mean-field theory (DFT + DMFT) has proven successful in describing strongly correlated materials and reproducing key spectroscopic features~\cite{PhysRevB.101.125123,PhysRevB.103.205134,PhysRevB.109.205132}. Early applications of DFT + DMFT implementations to Am, however, involved several simplifications. These include the lack of self-consistent total energy calculations~\cite{PhysRevLett.96.036404,PhysRevB.94.115148}, the use of the one-crossing approximation (OCA) for the quantum impurity problem~\cite{PhysRevB.64.155111}, and a reliance on idealized fcc lattices rather than experimentally determined crystal structures. Beyond Am, previous DFT + DMFT investigations have revealed quasiparticle multiplets in the Cm-III phase~\cite{PhysRevB.101.195123} and predicted a pressure-induced orbital-selective localized-itinerant transition in Cf, which is associated with valence state fluctuations~\cite{PhysRevB.99.045109}.

Despite substantial experimental and theoretical efforts, a unified physical picture of the electronic structure evolution across the Am, Cm, Bk and Cf series remains elusive~\cite{RevModPhys.98.015004}. Until now, the evolution of 5$f$ states across Am, Cm, Bk and Cf has been investigated using conventional density functional theory and various theoretical models. These studies have elucidated the critical role of $f-c$ hybridization and itinerant-localized nature of 5$f$ states in determining structural stability~\cite{McMahan19891}. Furthermore, high-pressure investigations on Cm, Bk and Cf have revealed pressure-induced delocalization of 5$f$ states, although the behavior of the still-localized 5$f$ states in Am-IV remains under debate~\cite{transPutheory2000,2006ActinideBonding,Johansson1995211}. However, while most existing theoretical work has focused on lighter actinides such as U, Np, and Pu~\cite{shim:2007,aPuZhu2013,dai:2003,SavrasovdPu2001}, systematic research dedicated to Am, Cm, Bk and Cf remains scarce. 

Several critical gaps persist. First, the continuous evolution of the 5$f$ electronic states with increasing electron count, encompassing the transition from localized to itinerant behavior, the evolution of strong electron correlation, and the change in angular momentum coupling schemes, has yet to be systematically elucidated. Second, the intrinsic relationship between 5$f$ bonding behavior and lattice stability under ambient-pressure double hexagonal close-packed (dhcp) and high-pressure face-centered cubic (fcc) structures remains insufficiently explored. These gaps are further underscored by experimental controversies that demand deep theoretical insight~\cite{PhysRevB.82.201103,heathman2000,PhysRevB.63.214101,EPL82.57007,PhysRevB.99.224419,Haire200763,Haire2006130}. The ambient-pressure dhcp and high-pressure fcc structures serve as two representative prototypes, dhcp represents the equilibrium structure at ambient conditions, while fcc is the common high-pressure phase shared by all four elements, enabling a systematic comparison of structural effects on 5$f$ electron correlations. In particular, the debated antiferromagnetic ground state of Cm and the pressure-induced valence state fluctuation observed in Cf call for a universal theoretical understanding grounded in the microscopic mechanisms of electron correlation.

This situation gives rise to several pressing questions. First, how do the correlation strength, degree of localization, orbital-selective correlation, electron occupancy, and angular momentum coupling of 5$f$ electrons systematically evolve with increasing atomic number from Am to Cf? Second, how does the electronic structure give rise to the diverse behaviors observed across this series, from the high-pressure superconductivity of Am, to the magnetic ground states of Cm and Bk, and to the valence state fluctuations in Cf? Third, how do the dhcp structure at ambient pressure and the fcc structure under pressure finely modulate the correlation strength and bonding behavior of 5$f$ electrons? Answering these questions is essential for constructing a unified physical picture that connects the microscopic electronic structure to the macroscopic properties of heavy actinides, for predicting their behavior under extreme conditions, and for understanding their material performance in actinide-related applications.

To address the above questions, we systematically investigate the electronic structures of Am, Cm, Bk, and Cf in both dhcp and fcc structures using a charge fully self-consistent DFT + DMFT approach~\cite{RevModPhys.68.13,RevModPhys.78.865,PhysRevB.81.195107}, implemented with the continuous-time hybridization expansion quantum Monte Carlo (CT-HYB) impurity solver. Key physical quantities including the density of states, band structures, hybridization functions, self-energy functions, valence state fluctuations, quasiparticle weights, effective electron masses, and x-ray branching ratios have been studied.
The remainder of this paper is organized as follows. Section~\ref{sec:method} outlines the computational framework of the DFT + DMFT methodology. In Section~\ref{sec:results}, we aim to delineate a complete physical picture of the continuous evolution of 5$f$ electron correlation strength from Am to Cf, and to reveal the orbital-selective correlation between the $j$ = 5/2 and $j$ = 7/2 electronic states. We further elucidate the modulation mechanisms of crystal structure (dhcp vs. fcc) on electron correlation strength, quantitatively characterize the strength of valence state fluctuations, and establish its intrinsic connection with the degree of 5$f$ electron localization. In addition, we analyze the angular momentum coupling schemes and the microscopic origins of magnetic ground states, thereby providing a theoretical foundation for understanding the competition between spin-orbit coupling and exchange interactions in transplutonium metals. Section~\ref{sec:discussion} discusses the kinetic energy, potential energy, spin susceptibility, and charge susceptibility for the dhcp and fcc phases to trace the progressive localization of 5$f$ electrons from Am to Cf. Finally, a concise summary is presented in Section~\ref{sec:summary}.

\section{Methods\label{sec:method}}
The density functional theory combined with the embedded dynamical mean-field theory (DFT + DMFT) is used to investigate the correlated 5$f$ electrons of transplutonium elements Am, Cm, Bk and Cf. This approach reformulates the many-body correlation problem as a self-consistent quantum impurity model, enabling an accurate treatment of strong Coulomb interactions while retaining the realistic band structure of the material~\cite{RevModPhys.68.13,RevModPhys.78.865}.

\textbf{DFT calculations.} DFT calculations were carried out using the \texttt{WIEN2K} code, which implements a full-potential linearized augmented plane-wave formalism~\cite{wien2k}. The experimental crystal structure of the dhcp and fcc phases were adopted for Am, Cm, Bk and Cf~\cite{heathman2000,Science.309.110,HAIRE1984119,PhysRevB.87.214111}. In these calculations, the muffin-tin sphere radii for these elements were set to 2.50 a.u., and the plane-wave cutoff parameter of $R_{\text{MT}}K_{\text{MAX}} = 8.0$. The valence states for each element consisted of the $5f$, $6d$, and $7s$ orbitals. The exchange-correlation potential was treated using the generalized gradient approximation (GGA) with the Perdew-Burke-Ernzerhof (PBE) functional~\cite{PhysRevLett.77.3865}, while spin-orbit coupling was included variationally. In accordance with high-pressure experimental results at 290~K, Am, Cm, Bk, and Cf were all treated as nonmagnetic.

\textbf{DFT + DMFT calculations.} We performed the DFT + DMFT calculations using the \texttt{eDMFT} software package~\cite{PhysRevB.81.195107}. The inverse temperature was set to $\beta$=40 in units of $(k_B\cdot T)^{-1}$, corresponding to $T\approx290$~K. The correlated nature of the $5f$ orbitals was treated within the DMFT formalism. The Coulomb interaction matrix for $5f$ orbitals was constructed using Slater integrals. For Am, we employed a Coulomb repulsion interaction parameter $U$=5.0~eV and Hund's exchange interaction parameter $J_{\text{H}}$=0.6~eV. For Cm, Bk and Cf, we used $U$=7.0~eV and $J_{\text{H}}$=0.6~eV. The local impurity Hamiltonian was formulated in the $|J, J_z\rangle$ basis. The DMFT projectors, which map the Kohn-Sham states onto the localized impurity basis, were constructed within a large energy window spanning from -10~eV to 10~eV relative to the Fermi level.

The constructed multi-orbital Anderson impurity models were solved using the hybridization expansion continuous-time quantum Monte Carlo impurity solver (dubbed as CT-HYB)~\cite{RevModPhys.83.349,PhysRevLett.97.076405,PhysRevB.75.155113}. This approach stochastically samples diagrams arising from the expansion of the partition function in powers of the impurity-bath hybridization. At each Monte Carlo step, the partition function is expressed as a sum over configurations of imaginary-time creation and annihilation operators, with the statistical weight of each configuration given by the determinant of a hybridization function matrix. The local impurity Green's function $G(i\omega_n)$ and self-energy $\Sigma(i\omega_n)$ are obtained by averaging over these configurations retaining fourteen 5$f$ orbitals. 
The DMFT self-consistency condition is then applied to update the local Green's function via $G^{-1}_{loc}(i\omega_n) = G^{-1}_{0}(i\omega_n)-\Sigma(i\omega_n)$, where $G_0$ is the bath Green's function, followed by an update of the hybridization function $\Delta(i\omega_n)$. The entire cycle is repeated until convergence is achieved.

To maintain computational tractability, the local Hilbert space was truncated by retaining only atomic eigenstates with electron occupancies $N \in [5,7]$ for Am, $N \in [6,9]$ for Cm, $N \in [7,10]$ for Bk, and $N \in [8,11]$ for Cf. A double-counting correction is required to avoid including electron–electron interactions both in the DFT exchange-correlation functional and in the explicitly added Hubbard term. Here we adopted the fully localized limit (FLL) scheme~\cite{jpcm:1997}, which is derived from the atomic limit and is widely considered appropriate for strongly correlated electron systems. The fully localized limit double-counting term is given by 
\begin{equation}
\Sigma_{dc} = U \left(n_{5f} - \frac{1}{2}\right) - \frac{J_{H}} {2} \left(n_{5f} -1 \right),
\end{equation}
where $n_{5f}=6.0, 7.0, 8.0, 9.0$ were chosen as the nominal occupancy of $5f$ orbitals for Am, Cm, Bk, and Cf, respectively, consistent with theoretical studies~\cite{PhysRevLett.96.036404,PhysRevB.101.195123,PhysRevB.99.045109}. Charge fully self-consistent DFT + DMFT calculations were performed, requiring approximately $60 \sim 80$ DFT + DMFT iterations to achieve good convergence. The convergence criteria for charge density and total energy were set to $10^{-4}$~e and $10^{-4}$~Ry, respectively. The final output were Matsubara self-energy function $\Sigma(i\omega_n)$ and impurity Green's function $G(i\omega_n)$, which were then utilized to obtain the integral spectral functions $A(\omega)$ and momentum-resolved spectral functions $A(\mathbf{k},\omega)$. In addition, analysis of the atomic eigenstate probabilities provided key insights into the 5$f$ valence state fluctuations and the underlying electronic configurations.

\section{Results\label{sec:results}}

\subsection*{A1. Quasiparticle resonance peaks in dhcp phase}

\begin{figure*}[!htb]
\includegraphics[width=\textwidth]{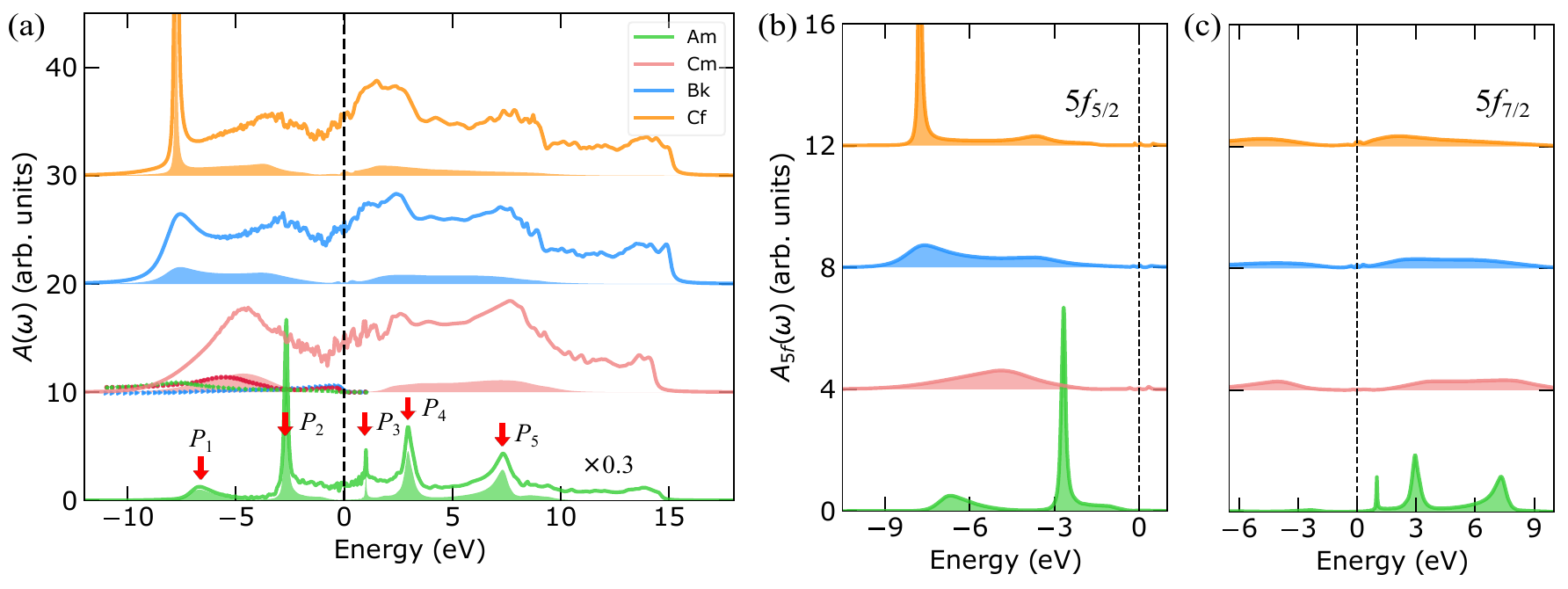}
\caption{(Color online) Total and 5$f$ partial density of states for dhcp phase of Am, Cm, Bk and Cf at $T = 290$~K by DFT + DMFT calculations. (a) Total density of states A($\omega$) (solid lines) and 5$f$ partial density of states A$_{5f}$($\omega$) (colored shaded regions). Experimental spectra for Cm-I are also presented. Blue right-pointing triangles and filled green five-pointed stars represent the valence-band photoemission spectra measured using He-I (21.22~eV) and He-II (40.81~eV) radiation, respectively. The red solid circles denote the 5$f$ contribution to the valence-band photoemission spectrum of Cm derived from He-II-$\beta$ (48.4~eV) radiation data. (b) and (c) Orbital-resolved (or $j$-resolved) 5$f$ partial density of states $j$ = 5/2 and $j$ = 7/2, respectively. The Fermi level $E_{F}$ is indicated by the vertical dashed lines. \label{fig:dosdhcp}}
\end{figure*}

The valence-band spectra of Cm-I, investigated using photoemission spectroscopy, are presented in Fig.~\ref{fig:dosdhcp}(a). Under He-I excitation, the spectrum is predominantly contributed by 6$d$ electrons. In contrast, both He-II and He-II-$\beta$ sources primarily detect the 5$f$ states, which manifest as a prominent feature around -5~eV~\cite{PhysRevB.83.125111}. Our calculations successfully reproduce this key experimental observation, capturing a distinct 5$f$-derived peak at around -5~eV, thereby further validating the reliability of our computational approach. A minor discrepancy in the precise peak position exists, which is mainly attributed to the inherent limitations of the single-site dynamical mean-field framework, specifically, the neglect of non-local spatial correlations. This neglect also leads to an underestimation of the spectral weight distribution at high energies and the full extent of lifetime broadening. Nevertheless, the reproduction of the characteristic 5$f$ peak confirms the effectiveness and robustness of our theoretical methodology.

The calculated total and partial density of states (DOS) of 5$f$ electrons for Am, Cm, Bk, and Cf in the double-hexagonal close-packed (dhcp) [Fig.~\ref{fig:dosdhcp}] and face-centered cubic (fcc) [Fig.~\ref{fig:dosfcc}] phase reveal a continuous evolution of 5$f$ states with increasing atomic number from Am to Cf. The density of states not only provides energy distribution of electrons but also offers insights into underlying properties such as conductivity, magnetism, correlation strength, orbital hybridization, and many-body quantum phenomena. The trend is vividly illustrated across the actinide series. Am displays localized 5$f$ electrons with well-separated spin-orbit-split subbands. Cm and Bk enter a regime characterized by distinct Hubbard bands driven by strong correlations, and Cf displays a sharp resonance peak signaling valence state fluctuations and Kondo-like many-body effects.

For the dhcp phase of Am, the total density of states [Fig.~\ref{fig:dosdhcp} (a)] exhibits substantial spectral weight near the Fermi level, characteristic of metallic behavior. However, the 5$f$ partial density of states around the Fermi level is nearly negligible, consistent with a non-magnetic ground state of total angular momentum $J$ = 0. The spectrum is dominated by five quasiparticle resonance peaks located at $P_1$ (-7.0~eV), $P_2$ (-2.8~eV), $P_3$ (1.0~eV), $P_4$ (3.0~eV), and $P_5$ (7.0~eV). As shown in Fig.~\ref{fig:dosdhcp} (b) and (c), the 5$f$ states are split by spin-orbit coupling into $j$ = 5/2 and $j$ = 7/2 manifolds, located predominantly below and above the Fermi level, respectively. The 5$f$ electrons are highly localized, manifesting as a sharp peak ($P_2$) that arises almost exclusively from the occupied $j$ = 5/2 states, corresponding to the 5$f^{6} (^{7}F_0)\rightarrow$ 5$f^{5} (^{6}H_{5/2})$ transition. Above the Fermi level, three weaker peaks ($P_3$, $P_4$ and $P_5$) originate mainly from unoccupied $j$ = 7/2 states. Peaks $P_3$ and $P_4$ are associated with the transitions 5$f^{6} (^{7}F_0)\rightarrow$ 5$f^{7} (^{8}S_{7/2})$ and 5$f^{6} (^{7}F_0)\rightarrow$ 5$f^{7} (^{6}P_{7/2})$, respectively.

The electronic structure of the dhcp phase of Cm undergoes a fundamental change. The density of states demonstrates a well-defined upper and lower Hubbard band structure: the lower Hubbard band (LHB) spans the energy range from -6 to -3~eV, while the upper Hubbard band (UHB) lies between 3 and 9~eV. Orbital-resolved analysis reveals that the LHB is primarily composed of $j$ = 5/2 states, whereas the UHB originates mainly from $j$ = 7/2 states. This contrasts sharply with the quasiparticle peaks features observed in the dhcp phase of Am, indicating that the 5$f$ electrons are strongly localized. For the dhcp phase of Bk, the overall spectral features resemble those of Cm, with a gapped Hubbard band structure. Nevertheless, the Hubbard bands shift systematically to lower and higher energies. The LHB moves to energy range of -8 to -3.5~eV, and the UHB appears between 2 and 7~eV. This progressive band narrowing, marked by the shift of Hubbard bands toward more negative energies, reflects the strengthening of electron correlations with increasing atomic number.

Unlike the Hubbard bands in Cm and Bk, this sharp resonance appears at a significantly lower binding energy, indicating a distinct electronic excitation mechanism. The total density of states presents a remarkably sharp resonance peak near -7.7~eV, accompanied by a pronounced transfer of spectral weight from the UHB toward the LHB. Orbital decomposition attributes this sharp feature and the transferred weight primarily to the $j$ = 5/2 states. This behavior is a hallmark of valence state fluctuations in strongly correlated systems. The sharp resonance corresponds to a heavy quasiparticle state arising from a Kondo regime, suggesting that the ground state of Cf lies in a region where strong correlations coexist with moderate valence state fluctuations, distinct from the fully localized 5$f$ states observed in Cm and Bk.

\subsection*{A2. Quasiparticle resonance peaks in fcc phase}

\begin{figure*}[!htb]
\includegraphics[width=\textwidth]{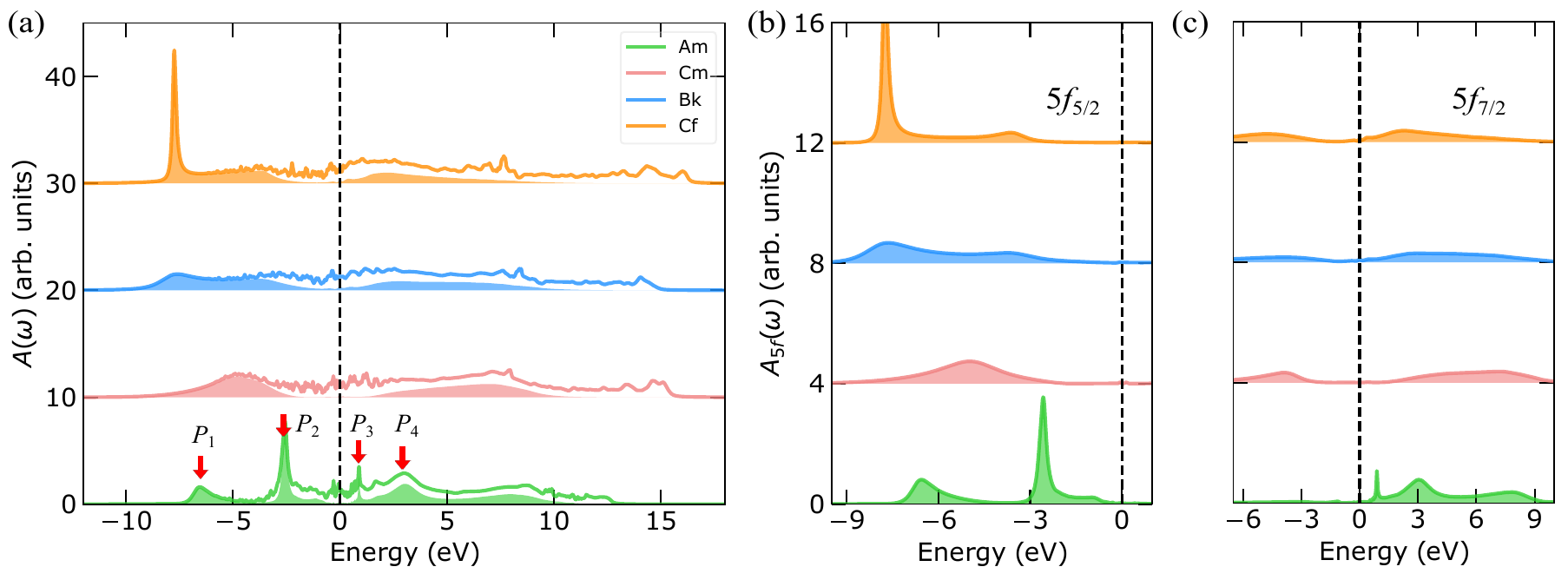}
\caption{(Color online) Total and 5$f$ partial density of states for fcc phase of Am, Cm, Bk and Cf at $T = 290$~K by DFT + DMFT calculations. (a) Total density of states A($\omega$) (solid lines) and 5$f$ partial density of states A$_{5f}$($\omega$) (colored shaded regions). (b) and (c) Orbital-resolved (or $j$-resolved) 5$f$ partial density of states $j$ = 5/2 and $j$ = 7/2, respectively. The Fermi level $E_{F}$ is indicated by the vertical dashed lines. \label{fig:dosfcc}}
\end{figure*}

For the fcc phase of Am [Fig.~\ref{fig:dosfcc} (a)], the main peak $P_2$ of total density of states shifts slightly toward the Fermi level to –2.6~eV, accompanied by a gradual transfer of spectral weight to energies closer to the Fermi level. The 5$f$ electrons remain predominantly localized. Additionally, a finite total density of states at the Fermi level confirms metallic behavior. Similar to the dhcp phase of Am, $P_3$ and $P_4$ peaks emerge above the Fermi level around 0.9 and 3.0~eV, again attributed to unoccupied $j$ = 7/2 states, with the $P_4$ peak exhibiting a broader spectral envelope.

For fcc phase of Cm, a profound change in the electronic structure is observed. The total density of states is characterized by well-separated lower and upper Hubbard bands. The LHB spans from -6 to -4~eV, while the UHB lies between 3 and 9~eV. The vanishing spectral weight at the Fermi level signals the presence of localized 5$f$ states. The LHB is dominated by $j$ = 5/2 states with only minor $j$ = 7/2 admixture, whereas the UHB arises primarily from $j$ = 7/2 states. Compared to the dhcp phase, the Hubbard bands in the fcc phase of Cm shift to higher energies and exhibit a wider bandgap, indicating a modulation of electron correlation strength by crystal structure.

For fcc phase of Bk, the overall density of states resembles that of fcc phase of Cm, but with the Hubbard bands shifted systematically to lower energies. The LHB moves downward to the range of -8 to -6~eV, and the UHB appears between 2 and 7~eV. Relative to the fcc phase of Cm, the LHB is lowered by approximately 2~eV, accompanied by a further widening of the bandgap. The orbital character remains largely unchanged, with the LHB primarily composed of $j$ = 5/2 states and the UHB derived mainly from $j$ = 7/2 states. In comparison with the dhcp phase, the fcc phase of Bk exhibits Hubbard bands shifted to lower energies and a broader gap, reflecting enhanced electron correlations in the fcc structure.

For fcc phase of Cf, the density of states displays evidently different features from those of Cm and Bk. An exceptionally sharp, high-intensity resonance peak emerges near -7.7~eV, accompanied by a weak shoulder around -3.0~eV. The UHB spans the range of 1.5 to 5.0~eV, with considerable spectral weight transferred from the UHB to the LHB region. Orbital-resolved analysis attributes this sharp resonance and the transferred weight predominantly to the $j$ = 5/2 states. 
A comparative analysis of the fcc and dhcp phases reveals that while the evolution of 5$f$ electronic behavior with atomic number follows a parallel trajectory, transitioning from metallic behavior (Am) to strongly localized 5$f$ states (Cm, Bk and Cf), the fcc phase consistently exhibits stronger correlation effects. Specifically, in the fcc structure, Cm, Bk and Cf manifest wider Hubbard bandgaps and more pronounced shifts of the Hubbard bands. These findings underscore that crystal structure plays a critical role in modulating 5$f$ electron localization, with the fcc phase favoring enhanced correlation strength compared to the dhcp phase.

\subsection*{B. Band structure} 

\begin{figure*}[!htb]
\includegraphics[width=0.9\textwidth]{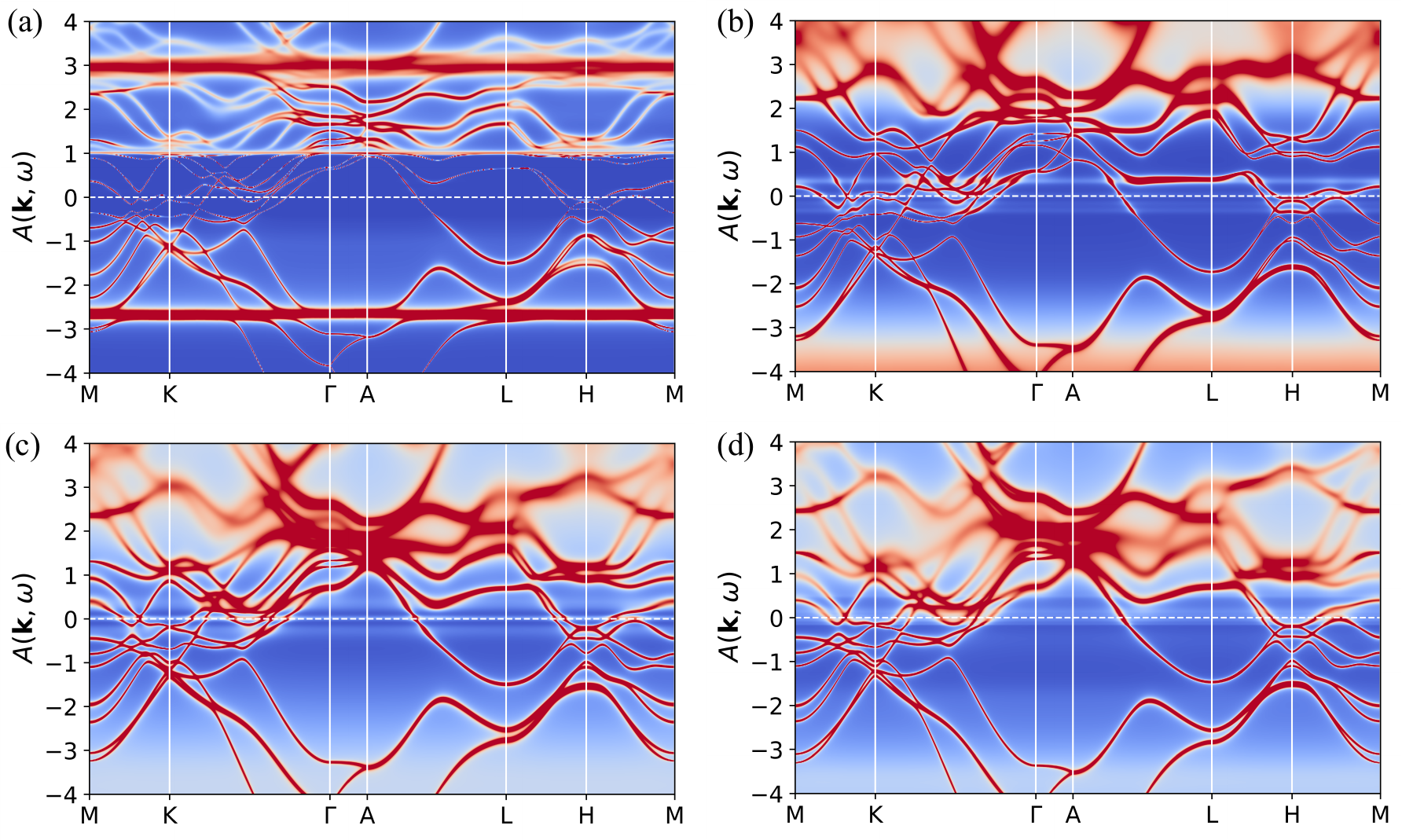}
\caption{(Color online). Momentum-resolved spectral functions A(${\bf k}$, $\omega$) of dhcp phase for Am, Cm, Bk and Cf at $T = 290$~K by the DFT + DMFT method. (a) Am. (b) Cm. (c) Bk. (d) Cf. The horizontal dashed lines denote the Fermi levels.
\label{fig:akwdhcp}}
\end{figure*}

\begin{figure*}[!htb]
\includegraphics[width=0.9\textwidth]{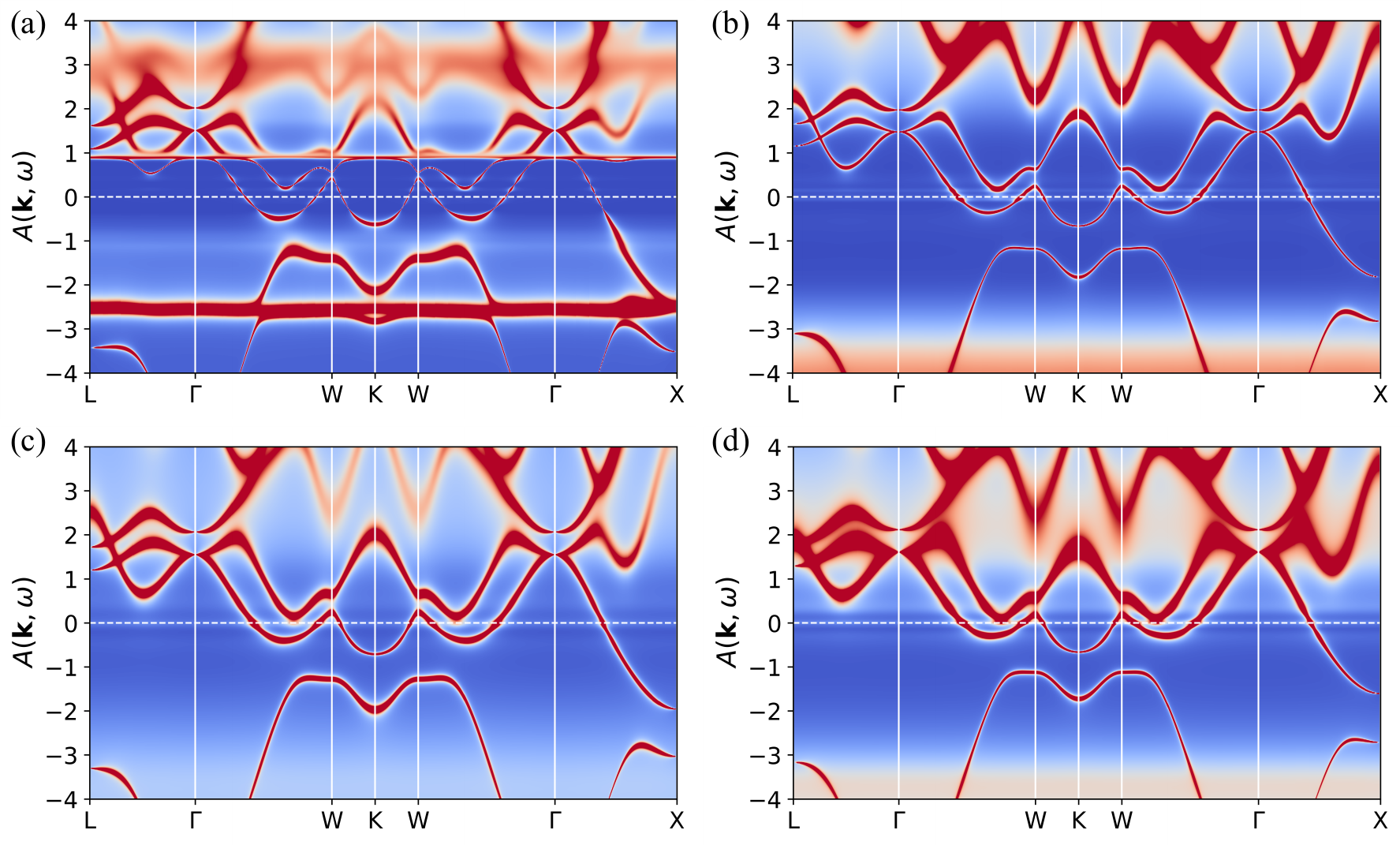}
\caption{(Color online). Momentum-resolved spectral functions A(${\bf k}$, $\omega$) of fcc phase for Am, Cm, Bk and Cf at $T = 290$~K by the DFT + DMFT method. (a) Am. (b) Cm. (c) Bk. (d) Cf. The horizontal dashed lines denote the Fermi levels.
\label{fig:akwfcc}}
\end{figure*}

The evolution of the electronic band structure along high-symmetry paths in the Brillouin zone accords with the calculated density of states. Figures~\ref{fig:akwdhcp}(a)–(d) illustrate the electronic bands along high-symmetry $k$-points for dhcp phase of Am, Cm, Bk, and Cf, respectively. Among these, the band structure of Am stands out as markedly different from the other three elements. Near the Fermi level, the spectral weight of the 5$f$ electrons is minimal; instead, it is mainly found around –2.8~eV below the Fermi level, as well as near 0.9 and 3.0~eV above it. Owing to $c$-$f$ hybridization, the lower and upper Hubbard bands broaden into relatively flat humps, reflecting the many-body character of the electronic states. With increasing pressure, the system stabilizes in the fcc structure [Fig.~\ref{fig:akwfcc}(a)–(d)]. Our calculated band shapes align closely with the earlier study~\cite{PhysRevLett.96.036404}, indicating that the 5$f$ bands continue to be concentrated around –2.8~eV below the Fermi level and near 1.0 and 3.0~eV above it. Even under moderate pressure, the 5$f$ electrons stay far from the Fermi level, preserving their strongly localized nature.

In the dhcp phase, Cm, Bk, and Cf exhibit highly similar electronic band structures. All three present characteristic features: tiny 5$f$ electrons spectral weight near the Fermi level, with lower and upper Hubbard bands forming below and above the Fermi level, respectively, signaling strong 5$f$ electron localization. Despite the overall similarity in band profiles, subtle differences exist in the spectral weight distribution of the conduction bands. Specifically, Cm exhibits relatively high conduction band spectral weight at approximately 2~eV above the Fermi level, whereas in Bk and Cf, the conduction band weight is primarily distributed across the energy range of 1 to 4~eV above the Fermi level.

The the fcc phase, Cm, Bk, and Cf share a highly similar electronic band structure, all exhibiting the strongly correlated 5$f$ states. In each case, the 5$f$ spectral weight vanishes at the Fermi level, with well-defined lower and upper Hubbard bands forming on either side, which is clear evidence of strong localization of 5$f$ electron. Despite this overarching similarity, subtle variations emerge in the conduction band spectral weight distribution. In Cm, the conduction band weight is concentrated in a relatively narrow feature around 2~eV above the Fermi level. In contrast, for Bk and Cf, this weight is more broadly distributed across the energy range from 1 and 4~eV above the Fermi level.

\subsection*{C. Hybridization function}

\begin{figure*}[!htb]
\includegraphics[width=0.9\textwidth]{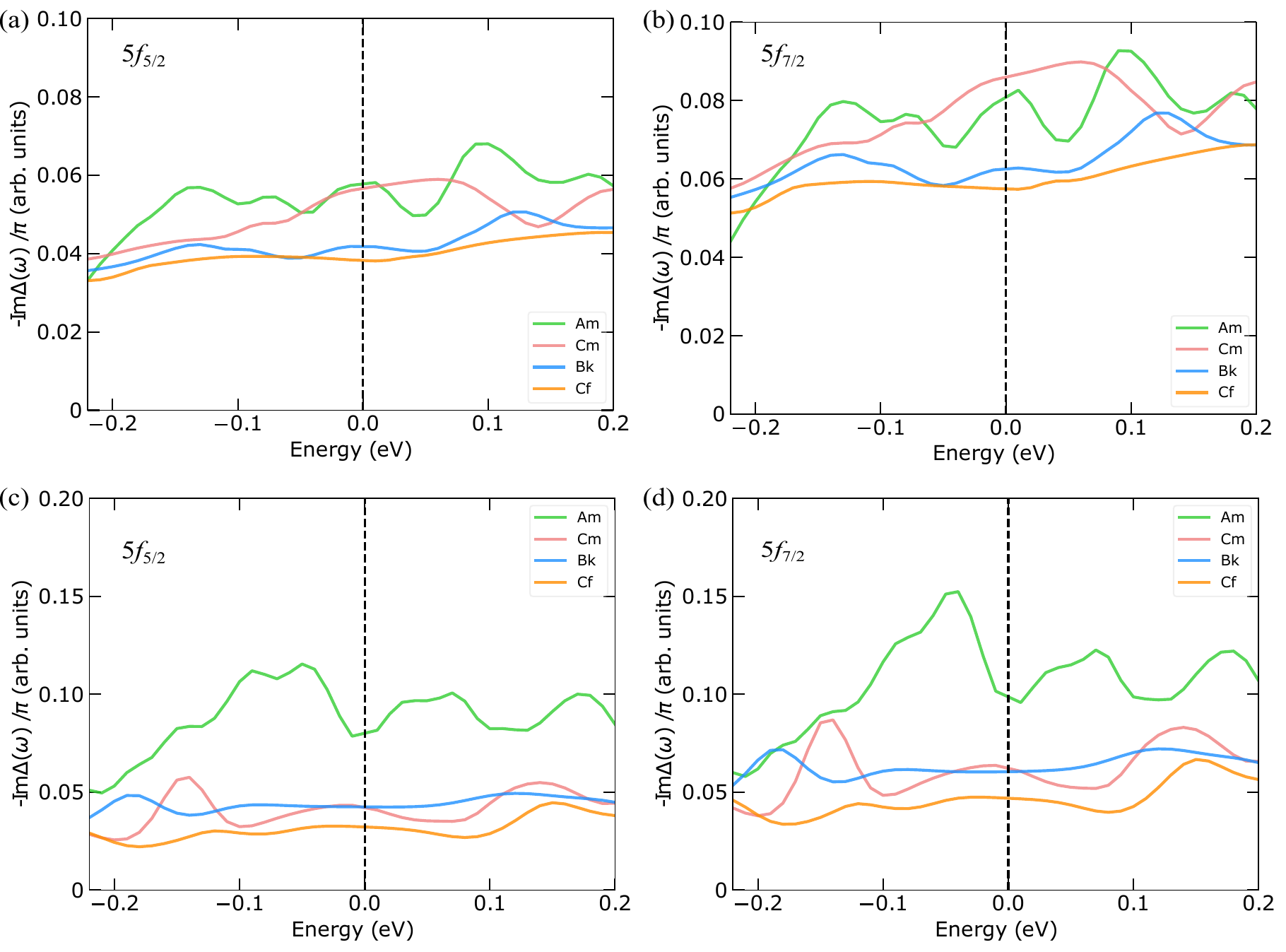}
\caption{(Color online) Hybridization functions of dhcp and fcc phases for Am, Cm, Bk and Cf at $T = 290$~K by the DFT + DMFT method. (a) For the 5$f$ $j$ = 5/2 states of dhcp phase. (b) For the 5$f$ $j$ = 7/2 states of dhcp phase. (c) For the 5$f$ $j$ = 5/2 states of fcc phase. (d) For the 5$f$ $j$ = 7/2 states of fcc phase. \label{fig:hyb}}
\end{figure*}

The hybridization function $\Delta(\omega)$ describes the coupling strength between localized $5f$ electrons and the surrounding conduction electron bath, serving as a key quantity for understanding charge fluctuations and itinerancy in correlated systems. We evaluated the impurity hybridization function $\tilde{\Delta}(\omega)$ for the 5$f$ orbitals as follows:
\begin{equation}
\label{eq:rehybd}
\tilde{\Delta}(\omega)=-\frac{\texttt{Im} \Delta(\omega)}{\pi}.
\end{equation}

Figure~\ref{fig:hyb} presents the impurity hybridization functions for the $5f_{5/2}$ and $5f_{7/2}$ states for the dhcp and fcc phases. For the $j = 5/2$ states [Fig.~\ref{fig:hyb}(a)], the hybridization intensity near the Fermi level follows the sequence Am $\approx$ Cm > Bk > Cf. Meanwhile, for the $j$ = 7/2 states [Fig.~\ref{fig:hyb}(b)], the order becomes to Cm > Am > Bk > Cf. Notably, Cm exhibits stronger hybridization than Am in the $j$ = 7/2 states, suggesting an orbital-dependent coupling to conduction electrons. This asymmetry may originate from the dominant $j$ = 7/2 states in the upper Hubbard band and their associated many-body effects.

For the fcc phase [Fig.~\ref{fig:hyb}(c) and (d)], pressure strikingly affects the hybridization strength. In Am, the hybridization intensities of both $j$ = 5/2 and $j$ = 7/2 states increase considerably compared to the dhcp phase, consistent with the expected pressure-induced enhancement of orbital overlap and band broadening. For Cm, Bk, and Cf, the relative order of hybridization intensity in the $j$ = 5/2 states and $j$ = 7/2 states remains Am > Cm $\approx$ Bk > Cf, but the absolute values change to varying degrees relative to the dhcp phase. In particular, Cm shows reduced hybridization in both $j$ = 5/2 and $j$ = 7/2 states under pressure. Across both dhcp and fcc phases, the hybridization intensity near the Fermi level is generally stronger for the $j$ = 7/2 states than for the $j$ = 5/2 states. This demonstrates that $j$ = 7/2 electrons couple strongly to the conduction electrons and may therefore be more susceptible to charge fluctuations, offering insight into the orbital-selective behavior observed later in the self-energy functions.

\subsection*{D. Atomic eigenstate probabilities}

\begin{figure*}[!htb]
\includegraphics[width=0.9\textwidth]{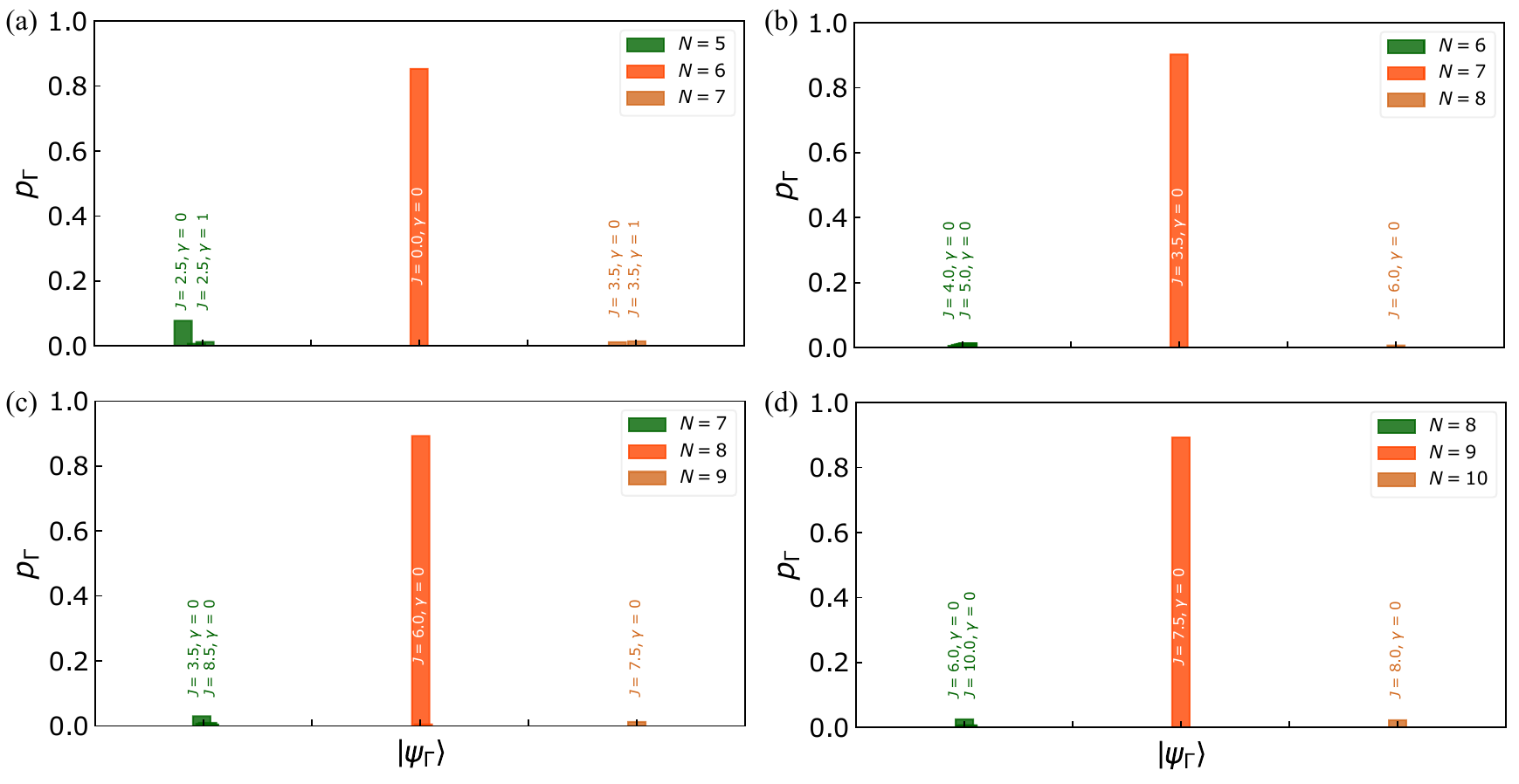}
\caption{(Color online) (a)–(d) Valence state histograms of dhcp phase for Am, Cm, Bk and Cf at $T = 290$~K by the DFT + DMFT method. The atomic eigenstates are denoted by using good quantum numbers $N$ (total occupancy), $J$ (total angular momentum) and $\gamma$ (the rest of the good quantum numbers), i.e., $|\psi_\Gamma\rangle=|N, J, \gamma \rangle$. The distributions of atomic eigenstate probabilities with respect to different $N$ are displayed in the legends. \label{fig:probdhcp}}
\end{figure*}

\begin{figure*}[!htb]
\includegraphics[width=0.9\textwidth]{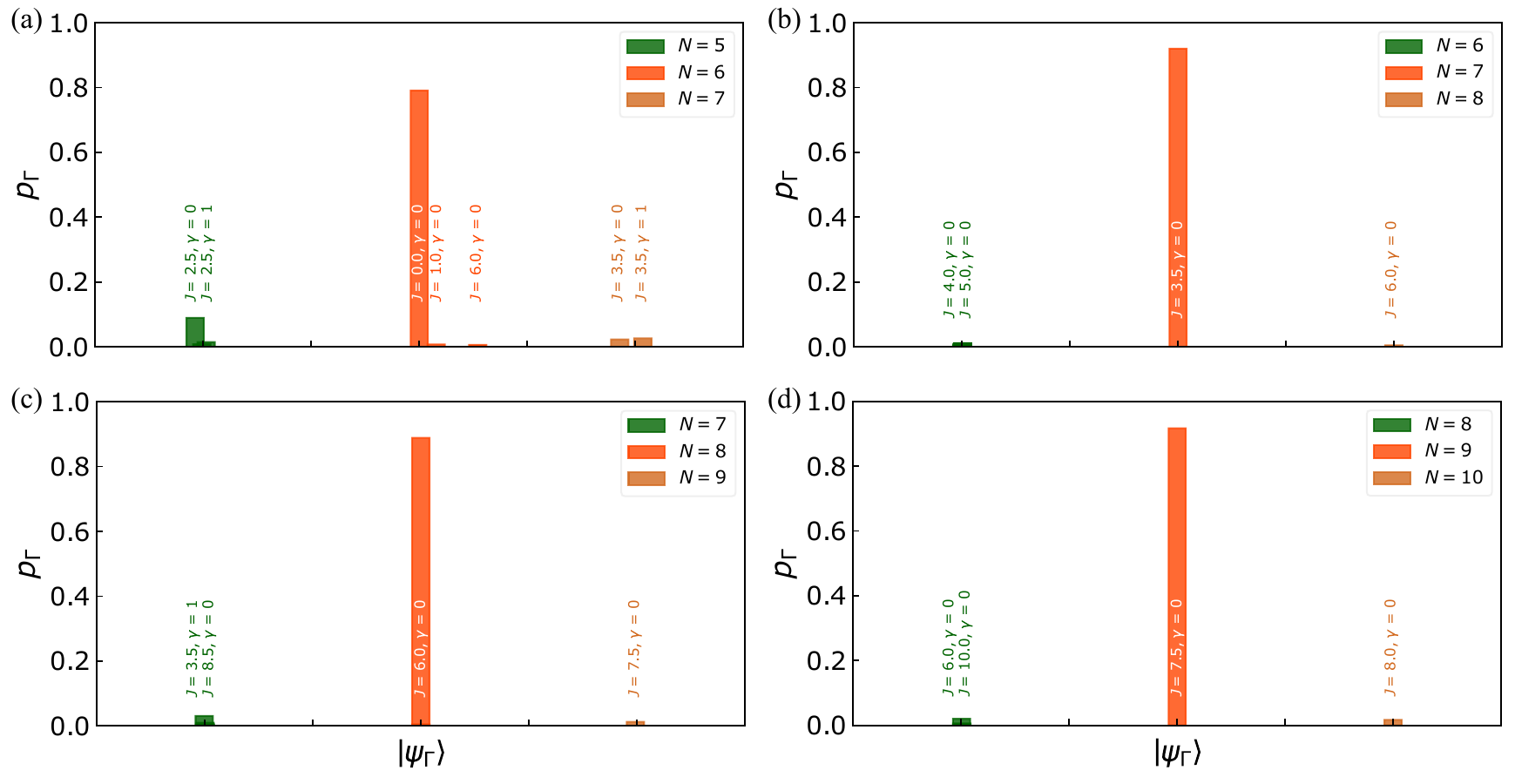}
\caption{(Color online) (a)–(d) Valence state histograms of fcc phase for Am, Cm, Bk and Cf at $T = 290$~K by the DFT + DMFT method. The atomic eigenstates are denoted by using good quantum numbers $N$ (total occupancy), $J$ (total angular momentum) and $\gamma$ (the rest of the good quantum numbers), i.e., $|\psi_\Gamma\rangle=|N, J, \gamma \rangle$. The distributions of atomic eigenstate probabilities with respect to different $N$ are displayed in the legends. \label{fig:probfcc}}
\end{figure*}

\begin{table*}[!htb]
\caption{The weights for 5$f$ electronic configurations ${w}(5f^n)$, 5$f$ orbital occupancy ($n_{5f}$, $n_{5/2}$ and $n_{7/2}$), total angular momentum $\langle J \rangle$, valence state fluctuation strength ${\cal V}$ and x-ray absorption branching ratio $\mathcal{B}$ for dhcp phase for Am, Cm, Bk and Cf at $T = 290$~K. \label{tab:probdhcp}}
\begin{center}
\begin{tabular}{cccccccccc}
\hline
\hline
Element & ${w}(5f^{n-1})$ & ${w}(5f^n)$ & ${w}(5f^{n+1})$ & $n_{5f}$ & $n_{5/2}$ & $n_{7/2}$ & $\langle J \rangle$ & ${\cal V}$ & $\mathcal{B}$ \\
\hline
Am ($n$=6) & 10.1\% $(5f^5)$ & 86.9\% $(5f^6)$ & 2.9\% $(5f^7)$ & 5.928 & 5.303 & 0.625 & 0.424 & 0.265 & 0.919 \\
Cm ($n$=7) & 7.5\% $(5f^6)$ & 90.8\% $(5f^7)$ & 1.6\% $(5f^8)$ & 6.941 & 4.357 & 2.584 & 3.525 & 0.184 & 0.783 \\
Bk ($n$=8) & 7.4\% $(5f^7)$ & 90.5\% $(5f^8)$ & 2.1\% $(5f^9)$ & 7.947 & 5.196 & 2.751 & 5.920 & 0.202 & 0.876 \\
Cf ($n$=9) & 6.5\% $(5f^8)$ & 90.5\% $(5f^9)$ & 3.0\% $(5f^{10})$ & 8.964 & 5.663 & 3.301 & 7.420 & 0.202 & 0.938 \\
\hline
\hline
\end{tabular}
\end{center}
\end{table*}

\begin{table*}[!htb]
\caption{The weights for 5$f$ electronic configurations ${w}(5f^n)$, 5$f$ orbital occupancy ($n_{5f}$, $n_{5/2}$ and $n_{7/2}$), total angular momentum $\langle J \rangle$, valence state fluctuation strength ${\cal V}$ and x-ray absorption branching ratio $\mathcal{B}$ for fcc phase for Am, Cm, Bk and Cf at $T = 290$~K. \label{tab:probfcc}}
\begin{center}
\begin{tabular}{cccccccccc}
\hline
\hline
Element & ${w}(5f^{n-1})$ & ${w}(5f^n)$ & ${w}(5f^{n+1})$ & $n_{5f}$ & $n_{5/2}$ & $n_{7/2}$ & $\langle J \rangle$ & ${\cal V}$ & $\mathcal{B}$ \\
\hline
Am ($n$=6) & 11.7\% $(5f^5)$ & 82.7\% $(5f^6)$ & 5.5\% $(5f^7)$ & 5.938 & 5.236 & 0.702 & 0.634 & 0.365 & 0.912 \\
Cm ($n$=7) & 6.4\% $(5f^6)$ & 92.4\% $(5f^7)$ & 1.2\% $(5f^8)$ & 6.947 & 4.351 & 2.596 & 3.518 & 0.153 & 0.782 \\
Bk ($n$=8) & 7.6\% $(5f^7)$ & 90.2\% $(5f^8)$ & 2.2\% $(5f^9)$ & 7.944 & 5.192 & 2.752 & 5.917 & 0.208 & 0.875 \\
Cf ($n$=9) & 5.3\% $(5f^8)$ & 92.5\% $(5f^9)$ & 2.2\% $(5f^{10})$ & 8.971 & 5.673 & 3.298 & 7.440 & 0.158 & 0.939 \\
\hline
\hline
\end{tabular}
\end{center}
\end{table*}

The atomic eigenstate probability, or equivalently valence state histogram, provides a powerful probe for investigating the valence state fluctuation in strongly correlated materials~\cite{shim:2007}. This quantity, denoted as $p_\Gamma$, represents the probability of finding the valence electrons in a particular atomic eigenstate $|\psi_\Gamma \rangle$, typically labeled by good quantum numbers such as electron occupancy $N$ or total angular momentum $J$. When only one or two atomic eigenstates dominate (i.e., exhibit high probabilities), the system experiences weak or confined valence state fluctuations~\cite{PhysRevB.99.045109}. Conversely, strong valence state fluctuations correspond to a broad distribution over many eigenstates without a dominant electronic configuration~\cite{Lawrence1981}.
To analyze the valence state fluctuations and mixed-valence behavior, we extract the 5$f$ electron atomic eigenstates from the output of DMFT ground states. Here $p_\Gamma$ measures the probability for $5f$ electrons to reside in a given atomic eigenstate $\Gamma$. The average $5f$ occupancy is given by $\langle n_{5f} \rangle = \sum_\Gamma p_\Gamma n_\Gamma$, where $n_\Gamma$ is the electron count in each atomic eigenstate $\Gamma$. Finally, the probability of a specified $5f^n$ configuration is defined as $w(5f^{n}) = \sum_\Gamma p_\Gamma \delta (n-n_\Gamma)$.

In the dhcp phase[Fig.~\ref{fig:probdhcp}(a)–(d)], the 5$f$ electronic configurations for Am, Cm, Bk, and Cf are dominated overwhelmingly by a single primary occupation number. These dominant electronic configurations correspond precisely to their respective trivalent ionic states: 5$f^6$ for Am$^{3+}$, 5$f^7$ for Cm$^{3+}$, 5$f^8$ for Bk$^{3+}$, and 5$f^9$ for Cf$^{3+}$. Nevertheless, the width of these distributions and the resulting valence state fluctuations differ dramatically across the series. As shown in Table~\ref{tab:probdhcp}, although the 5$f^6$ electronic configuration is dominant (86.9\%) in Am, it is accompanied by substantial admixtures of $5f^5$ (10.1\%) and $5f^7$ (2.9\%), indicating moderate valence state fluctuations. In contrast, the electronic configurations of Cm, Bk, and Cf are remarkably localized. The probability of the primary electronic configuration surpasses 90\% in each case 90.8\% for Cm (5$f^7$), 90.5\% for Bk (5$f^8$), and 90.5\% for Cf (5$f^9$), with all other configurations collectively accounting for less than 10\%. This points to a nearly singular and well-defined electronic configuration. 

Accordingly, the valence state fluctuation strength parameter is defined as ${\cal V} = \sum_n P_n (1 - P_n)$, where $P_n$ denotes the occupancy number of 5$f$ electrons. Am exhibits a substantially larger ${\cal V}$ value of 0.265, signaling a tendency toward mixed-valence behavior. The lower and mutually similar ${\cal V}$ values for Cm (0.184), Bk (0.202), and Cf (0.202) reflect well-defined 5$f$ occupation numbers, minimal valence state fluctuations, and nearly localized 5$f$ states. Moreover, the average total angular momentum $\langle J \rangle$ increases progressively with the 5$f$ electron count, consistent with the growing orbital and spin magnetic moments.

Thus Am occupies a pivotal position at the threshold where 5$f$ electron behavior in the actinide series transitions from itinerant to localized. Its 5$f$ wave functions retain sufficient spatial extent for weak hybridization with neighboring atoms, fostering the observed configuration mixing and valence state fluctuations. By contrast, Cm epitomizes localization, its 5$f$ electrons being tightly bound by the exceptional quantum mechanical stability of the half-filled 5$f^7$ shell. For the heavier Bk and Cf, the actinide contraction draws the 5$f$ orbitals closer to the nucleus, and the enhanced Coulomb potential further deepens their localization, cementing their stable, trivalent ground states. Hence, while Cm, Bk, and Cf are archetypes of localized 5$f$ electron systems, Am stands apart. Its residual 5$f$ itinerancy marks it as a unique demarcation point in the electronic structure and physical properties of the actinide elements.

In the fcc phase[Fig.~\ref{fig:probfcc}(a)–(d)], the 5$f$ electron configuration distributions of Am, Cm, Bk, and Cf remain dominated by a single primary configuration, which continues to correspond to their respective trivalent ionic states: 5$f^6$ for Am$^{3+}$, 5$f^7$ for Cm$^{3+}$, 5$f^8$ for Bk$^{3+}$, and 5$f^9$ for Cf$^{3+}$. However, compared to the dhcp phase, the width of these distributions and the associated valence state fluctuations exhibit notable changes. As listed in Table~\ref{tab:probfcc}, although the 5$f^6$ electronic configuration remains leading in Am (82.7\%), the admixtures of $5f^5$ (11.7\%) and $5f^7$ (5.5\%) have increased relative to the dhcp phase, indicating more pronounced valence mixing. By contrast, the distributions for Cm, Bk, and Cf remain highly concentrated, with the probability of the dominant electronic configuration reaching 92.4\% for Cm's $5f^7$, 90.2\% for Bk's $5f^8$, and 92.5\% for Cf's $5f^9$. The remaining configurations (such as Cm's 5$f^6$ and 5$f^8$, Bk's 5$f^7$ and 5$f^9$, and Cf's 5$f^8$ and 5$f^{10}$) each account for a minimal proportion, reflecting nearly a single electronic configuration.

From the perspective of valence state fluctuations, the ${\cal V}$ value for Am in the fcc phase (0.365) is higher than in the dhcp phase (0.265), implying that the itinerancy of its 5$f$ electrons is further enhanced in the fcc structure, allowing for more active fluctuations between different valence states. This increase is directly correlated with the larger proportions of the $5f^5$ and $5f^7$ electronic configurations in the fcc phase, highlighting the significant influence of crystal structure on electronic behavior. In comparison, the ${\cal V}$ value for Cm in the fcc phase (0.153) is lower than in the dhcp phase (0.184), demonstrating stronger 5$f$ electron localization. The ${\cal V}$ value for Bk (0.208) remains essentially unchanged relative to the dhcp phase (0.202), showing only a slight increase. The ${\cal V}$ value for Cf (0.158) is saliently lower than in the dhcp phase (0.202), exhibiting a more stable electronic configuration. Overall, the ${\cal V}$ values for Cm, Bk, and Cf in the fcc phase remain at low levels, signifying that their 5$f$ electron occupation numbers are relatively fixed, valence state fluctuations are minimal, and they display highly localized states. Moreover, the trend of increasing average total angular momentum $\langle J \rangle$ with increasing 5$f$ electron count is consistently maintained in the fcc phase, in accordance with the physical principles governing orbital and spin magnetic moments.

In the fcc phase, Am remains situated in the critical region of the actinide series, where 5$f$ electrons transition from itinerant to localized behavior. Even though the valence state fluctuations in fcc phase of Am  are enhanced relative to the dhcp phase, the 5$f$ electrons remain predominantly localized. Cm, due to the exceptional stability of its half-filled 5$f^7$ shell, demonstrates even stronger electron localization in the fcc phase (as evidenced by its reduced ${\cal V}$ value), with its 5$f$ electrons being tightly bound to the nucleus. Similarly, the localization of Bk's 5$f$ electrons remains largely stable in the fcc phase, with only minor valence state fluctuations. For Cf, the contraction of the 5$f$ orbitals due to increased atomic number (actinide contraction) is fully realized in the fcc phase, where the Coulomb potential further localizes the electrons, thereby yielding a more stable electronic configuration than in the dhcp phase. Consequently, Cm, Bk, and Cf all exhibit distinctly localized 5$f$ character in the fcc phase. Nevertheless, valence state fluctuation strength in Am is enhanced in the fcc phase compared to the dhcp phase, further solidifying its unique position as a demarcation point in the electronic behavior of the actinide elements.

\subsection*{E. Self-energy functions}

\begin{figure*}[!htb]
\includegraphics[width=0.9\textwidth]{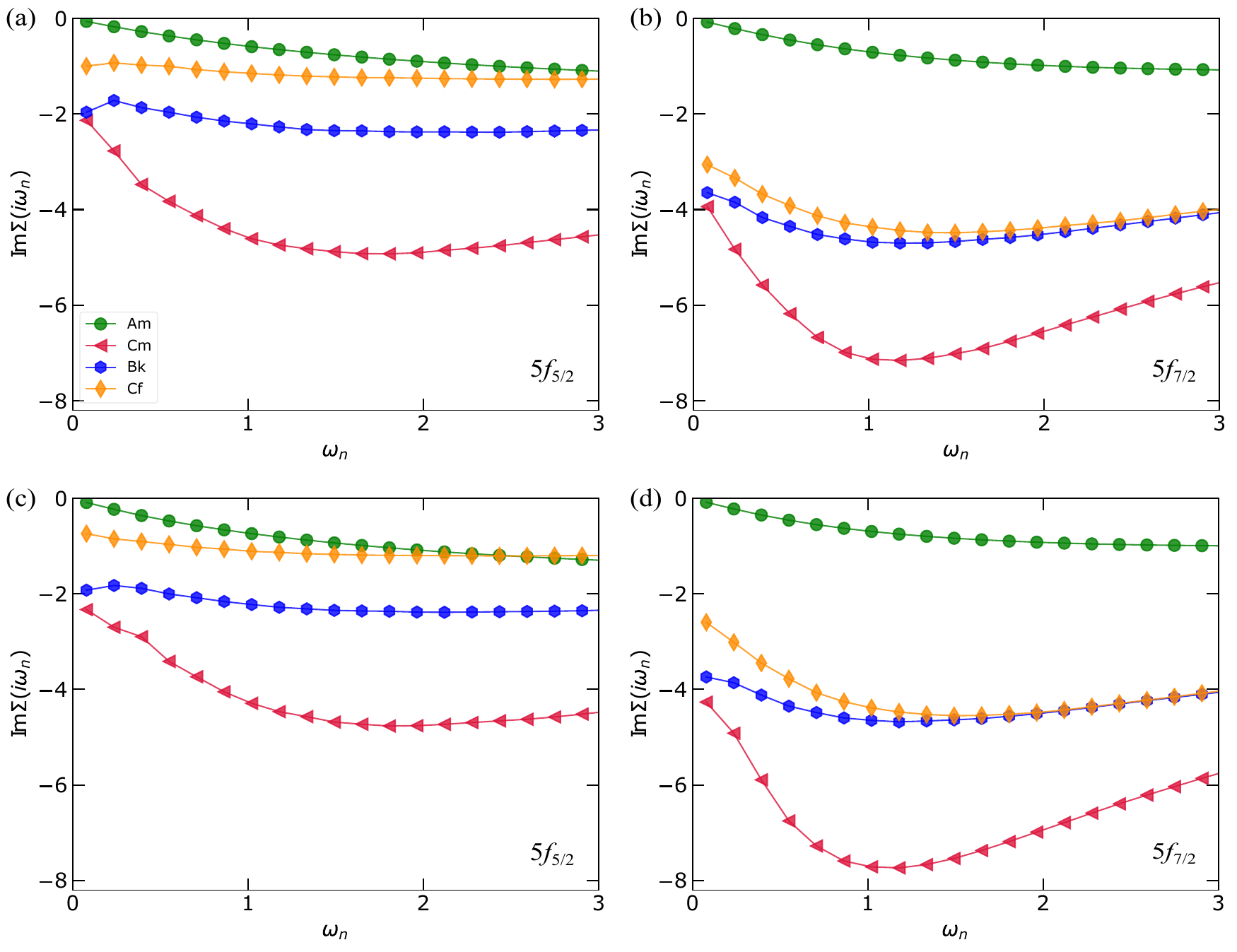}
\caption{(Color online) Imaginary parts of the Matsubara self-energy functions of Am, Cm, Bk and Cf at $T = 290$~K in the low-frequency regime by DFT + DMFT calculations. (a) For the 5$f$ $j$ = 5/2 states of dhcp phase. (b) For the 5$f$ $j$ = 7/2 states of dhcp phase. (c) For the 5$f$ $j$ = 5/2 states of fcc phase. (d) For the 5$f$ $j$ = 7/2 states of fcc phase.\label{fig:sig}}
\end{figure*}

\begin{table}[th]
\caption{Calculated orbital-dependent quasiparticle weights $Z$, effective electron masses $m^{*}$ and fitting parameter $\alpha$ for dhcp and fcc phases of Am, Cm, Bk and Cf at $T = 290$~K. \label{tab:sig}}
\begin{center}
\begin{tabular}{cccccccc}
\hline
\hline
Element & $Z_{5/2}$ & $Z_{7/2}$ & $m^{*}_{5/2}$ & $m^{*}_{7/2}$ & $R$ & $\alpha_{5/2}$ & $\alpha_{7/2}$  \\
\hline
Am (dhcp) & 0.555 & 0.505 & 1.801$m_e$  & 1.980$m_e$  & 1.099 & 0.500 & 0.443 \\
Cm (dhcp) & 0.036 & 0.019 & 27.153$m_e$ & 51.133$m_e$ & 1.895 & 0.417 & 0.263 \\
Bk (dhcp) & 0.038 & 0.021 & 25.987$m_e$ & 47.422$m_e$ & 1.810 & 0.110 & 0.180 \\
Cf (dhcp) & 0.073 & 0.025 & 13.700$m_e$ & 39.950$m_e$ & 2.920 & 0.011 & 0.155 \\
\hline
Am (fcc) & 0.468 & 0.481 & 2.139$m_e$  & 2.079$m_e$  & 0.973 & 0.730 & 0.670 \\
Cm (fcc) & 0.033 & 0.018 & 30.682$m_e$ & 55.332$m_e$ & 1.833 & 0.120 & 0.140 \\
Bk (fcc) & 0.039 & 0.021 & 25.503$m_e$ & 48.627$m_e$ & 1.857 & 0.050 & 0.060 \\
Cf (fcc) & 0.096 & 0.029 & 10.443$m_e$ & 34.065$m_e$ & 3.310 & 0.030 & 0.320 \\
\hline
\hline
\end{tabular}
\end{center}
\end{table}

The self-energy functions encapsulate the strength of electron correlations. Figure~\ref{fig:sig} presents the imaginary part of the self-energy functions for the $j$ = 5/2 and $j$ = 7/2 states of Am, Cm, Bk, and Cf in both the dhcp and fcc phases. Notably, the behaviors in the two phases are highly similar across all elements, suggesting that the observed physical trends are intrinsic to the 5$f$ electron count and its inherent correlation strength, with the specific crystal structure (dhcp and fcc) playing only a minor and quantitative role. 

Americium serves as a moderately correlated reference point in the series. As shown in Fig.~\ref{fig:sig}, the imaginary parts of the Matsubara self-energy for the $j$ = 5/2 and $j$ = 7/2 states approach zero in the low-frequency limit, implying long-lived quasiparticles and good metallicity. This behavior is consistent across the dhcp and fcc structures. Furthermore, the self-energy curves are smooth and featureless, lacking the pronounced ``checkmark'' or ``upward convex'' curvature observed in the more correlated elements to follow.

The Cm system significantly deviates from Fermi liquid behavior, entering a strongly correlated metallic regime. Here, the LS coupling scheme prevails: Hund's rule in the half-filled 5$f^7$ shell forces a parallel spin alignment, generating local magnetic moments. This result corroborates with those previously reported by Moore et al.~\cite{PhysRevB.76.073105}. Consequently, in both dhcp and fcc phases, the imaginary parts of the Matsubara self-energy for both $j$ = 5/2 and $j$ = 7/2 states display distinct ``checkmark'' shapes. Crucially, these self-energies remain finite in the zero-frequency limit, indicating strong scattering between local moments and itinerant electrons that substantially suppresses quasiparticle lifetimes. Orbital differentiation is already evident at this stage. The absolute value of $\mathrm{Im}\Sigma(0)$ for the $j$ = 7/2 state is approximately twice that of the $j$ = 5/2 state. This disparity reveals that the $j$ = 7/2 electrons experience noticeably stronger scattering, leading to even shorter quasiparticle lifetimes and evincing the emergence of orbital-selective correlations in Cm.

The Bk system marks a significant escalation in both the strength of electron correlations and the degree of orbital-selective differentiation. Relative to the half-filled Cm configuration, the additional electron in Bk is forced into the $j$ = 7/2 orbital, thereby intensifying orbital polarization. This electronic configuration yields distinct orbital dynamics: in both dhcp and fcc structures, the imaginary part of the self-energy $\mathrm{Im}\Sigma(i\omega_n)$ for the $j$ = 5/2 state develops a slight ``upward convex'' curve at low frequencies, while the $j$ = 7/2 state exhibits a ``checkmark'' shape. This stark contrast identifies the $j$ = 7/2 orbital as the primary carrier of the local magnetic moments and the dominant source of localization. Both orbitals maintain finite zero-frequency self-energies, with the magnitude for the $j$ = 7/2 state consistently exceeding that of the $j$ = 5/2 state. This disparity confirms that the intense scattering of itinerant electrons originates predominantly from the $j$ = 7/2 orbital, establishing Bk as a prototypical metal with pronounced orbital-selective correlations.

Building on this trend, the Cf system represents an extreme culmination of this evolution. Remarkably, the angular momentum coupling scheme reverts to the $jj$ coupling framework. In both dhcp and fcc structures, the $j$ = 5/2 state retains a slight ``upward convex'' curve at low frequencies, qualitatively similar to Bk but with key quantitative differences discussed below. Concurrently, the $j$ = 7/2 state displays a sharp ``checkmark'' shape with zero-frequency absolute values much larger than those of $j$ = 5/2, establishing it as the epicenter of strong scattering. This configuration drives the system toward the brink of metallicity, underscoring the persistent and even intensified role of the $j$ = 7/2 subshell in dictating the correlated electron behavior.

In the low-frequency regime, the data reveal quasi-linear behavior, extrapolating to zero as $\omega_n \rightarrow$ 0. To quantify this trend, we employ the empirical formula:
\begin{equation}
\mathrm{Im}\Sigma(i\omega_n) = c(i\omega_n)^{\alpha} + b,
\end{equation}
where $b$, $c$, and $\alpha$ are fitting parameters. For Am, the extracted exponents $\alpha$ are approximately 0.5 in the dhcp phase and 0.7 in the fcc phase, deviating from the linear frequency dependence expected from Landau's Fermi-liquid theory~\cite{RevModPhys.68.13}. In Cm, the exponents $\alpha$ are 0.417 ($j$ = 5/2) and 0.263 ($j$ = 7/2) in the dhcp phase, and 0.120 ($j$ = 5/2) and 0.140 ($j$ = 7/2) in the fcc phase, all characteristic of non-Fermi liquid behavior. This trend continues in Bk, where $\alpha$ reaches 0.11 ($j$ = 5/2) and 0.18 ($j$ = 7/2) in the dhcp phase, and further drops to 0.05 ($j$ = 5/2) and 0.06 ($j$ = 7/2) in the fcc phase, again signaling a clear departure from Fermi-liquid ground states. In Cf, the exponents $\alpha$ for $j$ = 5/2 states approach zero (0.011 for dhcp phase and 0.030 for fcc phase), while those for $j$ = 7/2 states are 0.155 for dhcp phase and 0.320 for fcc phase. Both orbitals retain finite zero-frequency self-energies, and the near-vanishing exponents signify a complete breakdown of Fermi-liquid theory, placing the system at the very edge of metallicity.  
To further quantify the correlation strength, the self-energy function $\Sigma(\omega)$ is utilized to determine both the quasiparticle weight $Z$ and effective electron masses $m^{*}$ for Am-$5f$ electrons~\cite{RevModPhys.68.13}:
\begin{equation}
\label{eq:renor}
Z^{-1} = \frac{m^{*}}{m_e} = 1 - \frac{\partial \texttt{Im}\Sigma(i \omega_0)}{\partial \omega_0},
\end{equation}
where $\omega_0 \equiv \pi/\beta$ and $m_e$ is the bare electron mass. The calculated orbital-resolved quasiparticle weights $Z$ and effective electron masses $m^{*}$ for the dhcp and fcc phases of Am, Cm, Bk and Cf are summarized in Table~\ref{tab:sig}. These quantities provide a quantitative measure of the electron correlation strength and its orbital dependence across the series. To capture the orbital differentiation between the $5f_{5/2}$ and $5f_{7/2}$ manifolds, we define the ratio $R \equiv Z_{5/2}/Z_{7/2}$, where a value deviating from unity signifies the emergence of orbital-selective correlations.

In line with the self-energy analysis, Am resides in the moderate correlation regime. Its quasiparticle weights $Z$ range from 0.47 to 0.56, with corresponding effective electron masses $m^{*}$ of approximately 1.8 $m_e$ to 2.1 $m_e$, indicating moderate electron renormalization. The $R$ values are 1.099 for the dhcp phase and 0.973 for the fcc phase, both close to unity. This confirms that the $j$ = 5/2 and $j$ = 7/2 states experience comparable correlation strengths, with only a slight deviation from perfect orbital degeneracy. This behavior stands in contrast to the moderately orbital-selective correlations observed in the $5f$ electrons of plutonium, where a clear differentiation between the $5f_{5/2}$ and $5f_{7/2}$ states has been reported~\cite{PhysRevB.99.125113, PhysRevB.101.125123}.

Moving to Cm, the correlation strength increases dramatically. The quasiparticle weights $Z$ plummet to a range of 0.02 to 0.04, while the effective electron masses $m^{*}$ are correspondingly enhanced to 27 $m_e$ $\sim$ 55 $m_e$. This order-of-magnitude change signals a transition into the strongly correlated regime with pronounced electron localization. Furthermore, the $R$ values of 1.895 (dhcp) and 1.833 (fcc) now deviate significantly from unity, providing clear evidence that orbital-resolved correlation characteristics begin to emerge in Cm.

This trend of orbital differentiation intensifies in Bk, which exhibits even more pronounced inter-orbital disparity. Here, the quasiparticle weights are approximately 0.038 $\sim$ 0.039 for the $j$ = 5/2 state and 0.021 for the $j$ = 7/2 state, yielding effective electron masses $m^{*}$ of approximately 26 $m_e$ and 47 $m_e$ $\sim$ 48 $m_e$, respectively. The substantial difference between the two orbitals is quantified by the $R$ values of 1.810 (dhcp) and 1.857 (fcc), robustly confirming the establishment of orbital-selective correlations in Bk.

The Cf system represents the extreme endpoint of this evolutionary path. Despite the reversion to a $jj$ coupling scheme, its correlation strength reaches a critical point. The quasiparticle weights $Z$ are 0.073 $\sim$ 0.096 for the $j$ = 5/2 orbital but only 0.025 $\sim$ 0.029 for the $j$ = 7/2 orbital, corresponding to effective electron masses $m^{*}$ of 10 $\sim$ 14 $m_e$ and 34 $m_e$ $\sim$ 40 $m_e$, respectively. The inter-orbital disparity is the most pronounced in the series, with $R$ values soaring to 2.920 (dhcp) and 3.310 (fcc). This indicates that the $j$ = 7/2 electrons are approaching complete localization.

In summary, the dhcp and fcc phases exhibit highly consistent physical trends across all elements, with only minor quantitative differences in their $Z$, $m^{*}$, and $R$ values. This consistency confirms that the observed evolution from Am to Cf is an intrinsic property of the $5f$ electron system as a function of electron count. Cf reveals a counterintuitive physical picture: although its coupling scheme reverts to the $jj$ limit, the higher $5f$ electron count drives the system into a $jj$-framework quantum critical state through intra-orbital correlation effects within the $j$ = 7/2 subshell. This represents the ultimate outcome of the competition between metallicity and locality in the heavy actinide series.

\subsection*{F. Angular momentum coupling scheme}
As the 5$f$ electron count increases from Am to Cf, the occupancy gradually shifts between the $j$ = 5/2 and $j$ = 7/2 manifolds. This evolution determines the angular momentum coupling scheme and, consequently, influences chemical bonding and related physical properties. In multielectron systems, a competition arises between spin-orbit coupling and Coulomb interactions, leading to three distinct coupling regimes: Russell-Saunders (LS) coupling, $jj$ coupling, and intermediate coupling (IC)~\cite{RevModPhys.81.235}. Although intermediate coupling is often adopted to describe the ground states of late actinides~\cite{PuBx2022}, how these systems behave under high pressure remains an open question. Understanding the evolution of both 5$f$ occupancy and the associated angular momentum coupling in the high-pressure phases of late actinides is therefore essential, as it provides direct insight into the interplay between localization, itinerancy, magnetism, and relativistic effects under extreme compression.

Core-level spectroscopic techniques, such as electron energy-loss spectroscopy and x-ray absorption spectroscopy, offer a direct probe of the 5$f$ electronic states in actinides~\cite{PhysRevB.76.073105}. In both methods, a core electron is excited above the Fermi level, allowing access to the unoccupied density of states. When the excitation originates from the 4$d$ core level, the spin-orbit interaction per hole in the 5$f$ shell can be quantified using the spin-orbit sum rule. In such analyses, the branching ratio is extracted from the spin-orbit-split core edges, where dipole selection rules permit two types of transitions: $4d_{5/2} \rightarrow 5f_{5/2, 7/2}$ ($N_5$) and $4d_{3/2} \rightarrow 5f_{5/2}$ ($N_4$). Because an excited 4$d$ core electron can only populate specific 5$f$ final states, the resulting variation in the branching ratio provides a measurable signature that can be rigorously interpreted within the spin-orbit sum-rule formalism.

A key quantity derived from these measurements is the x-ray absorption branching ratio $\mathcal{B}$, defined as the relative intensity of the $4d_{5/2}$ absorption line~\cite{PhysRevA.38.1943}. This ratio serves as a direct experimental indicator of the spin-orbit coupling strength in the 5$f$ shell. Under the approximation that core-valence electrostatic interactions are neglected, $\mathcal{B}$ can be expressed analytically~\cite{shim:2007}. 

\begin{equation}
\label{eq:ratio}
\mathcal{B} = \frac{3}{5} - \frac{4}{15} \frac{1}{14 - n_{5/2} - n_{7/2}} \left ( \frac{3}{2} n_{7/2} - 2 n_{5/2} \right ),
\end{equation}
where $n_{5/2}$ and $n_{7/2}$ indicate the 5$f$ occupation numbers for the $5f_{5/2}$ and $5f_{7/2}$ states, respectively. The calculated results are summarized in Tables~\ref{tab:probdhcp} and~\ref{tab:probfcc}, where x-ray absorption branching ratio $\mathcal{B}$ exhibits a similar behavior for both the dhcp and fcc phases. So we illustrate here using the dhcp phase as a representative example. Our results are in excellent agreement with those previously reported by Moore et al.~\cite{PhysRevB.76.073105}. Their x-ray branching ratio of $\mathcal{B}$=0.93 corresponds to occupancies of $n_{5/2}$=5.38 and $n_{7/2}$=0.62, while our calculations yield $n_{5/2}$=5.303 and $n_{7/2}$=0.625, corresponding to $\mathcal{B}$=0.919.
The close consistency between the two datasets validates the reliability of our computational approach and provides strong evidence that the angular momentum coupling in Am approaches the $jj$ coupling limit. It is worth mentioning that the dominant occupancy of the $5f_{5/2}$ subshell and predominant spin-orbit coupling over exchange interaction drives the system toward the $jj$ coupling regime. Particularly, the EELS branching ratio is sensitive to the degree of 5$f$ electron delocalization, and the observed values reflect the localized nature of the 5$f$ electrons in Am. The x-ray absorption branching ratio can be experimentally probed using techniques such as electron energy-loss spectroscopy, x-ray absorption spectroscopy, and x-ray magnetic circular dichroism.

In late actinides, the interplay between spin-orbit coupling and exchange interaction generally leads to an intermediate coupling. For Cm, as the 5$f$ count increases from six to seven, a pronounced shift occurs in favor of the exchange interaction, marking the transition from optimal spin-orbit stabilization for 5$f^6$ to optimal exchange interaction stabilization for 5$f^7$. Thus, Cm achieves maximum energy stabilization through exchange interaction. This stabilization requires all spins to be parallel via Hund's rule coupling, a condition inherently satisfied in LS coupling. Consequently, Cm shifts toward the intermediate coupling regime with a strong bias toward LS coupling, as evidenced by x-ray absorption and x-ray magnetic circular dichroism measurements~\cite{PhysRevB.99.224419}.

This deviation results in a dramatic redistribution of 5$f$ electrons. Compared with Am, the dhcp phase of Cm transfers more than one electron into the $5f_{7/2}$ level, yielding calculated occupation numbers of $n_{5/2}$=4.357 and $n_{7/2}$=2.584. These values correspond to the x-ray branching ratio of $\mathcal{B}$=0.783, which aligns closely with previously reported values~\cite{PhysRevB.76.073105} ($\mathcal{B}$=0.794 corresponding to occupancies of $n_{5/2}$=4.41 and $n_{7/2}$=2.59). Additionally, Cm exhibits a modest orbital moment aligned parallel to the large spin moment, a consequence of nonvanishing spin–orbit coupling within the intermediate coupling regime. Together, these features elucidate the mechanism underlying the magnetic stabilization of Cm through the interplay between exchange interaction and spin-orbit coupling~\cite{PhysRevLett.98.236402}.

A similar trend persists in Bk, which has a nominal 5$f^8$ configuration. Bk falls almost directly in the intermediate coupling regime, shifted towards LS coupling, consistent with electron energy-loss spectroscopy measurements~\cite{Muller2021Cf}. In this regime, Hund's rule coupling requires all spins to be aligned in parallel, giving rise to a large spin moment, in agreement with previous results~\cite{RevModPhys.81.235}. In contrast, 5$f$ electrons in Cf are predominantly localized, leading to angular momentum coupling that approaches the $jj$ coupling limit. This indicates that spin-orbit coupling dominates over exchange interaction, in accordance with experimental findings~\cite{Muller2021Cf}. Thus, Am and Cf both approach the $jj$ coupling limit, where spin-orbit coupling prevails. Cm and Bk where exchange interaction dominates over spin-orbit coupling reside in an intermediate coupling regime with a strong bias toward LS coupling. This contrast reflects the competing influences of spin–orbit coupling and exchange stabilization, with the latter playing a more significant role in Cm and Bk~\cite{PhysRevB.38.3158}.

\section{Discussion\label{sec:discussion}}

\textbf{Evolution of energy scales and magnetic response with 5$f$ occupancy from Am to Cf.}

\begin{table}[th]
\caption{Calculated kinetic energy ($E_{\mathrm {kin}}$), potential energy ($E_{\mathrm {pot}}$), spin susceptibility ($\chi_{\mathrm {S0}}$) and charge susceptibility ($\chi_{\mathrm{D}}$) for dhcp and fcc phases of Am, Cm, Bk and Cf at $T = 290$~K. \label{tab:energy}}
\begin{center}
\begin{tabular}{cccccc}
\hline
\hline
Element & nominal 5$f$ count & $E_{\mathrm {kin}}$ (eV) & $E_{\mathrm {pot}}$ (eV) & $\chi_{\mathrm {S0}}$ & $\chi_{\mathrm{D}}$ \\
\hline
Am (dhcp) & 5$f^6$ & -0.86 & 66.72 & 0.296 & +0.01 \\
Cm (dhcp) & 5$f^7$ & -0.82 & 137.1 & 110.8 & -3.16 \\
Bk (dhcp) & 5$f^8$ & -0.71 & 186.5 & 553.6 & -1.30 \\
Cf (dhcp) & 5$f^9$ & -0.73 & 243.2 & 769.7 & -3.60 \\
\hline
Am (fcc) & 5$f^6$ & -1.24 & 67.21 & 0.543 & +0.01 \\
Cm (fcc) & 5$f^7$ & -0.62 & 137.2 & 203.6 & -0.29 \\
Bk (fcc) & 5$f^8$ & -0.78 & 186.4 & 354.9 & -4.92 \\
Cf (fcc) & 5$f^9$ & -0.57 & 243.5 & 811.1 & -4.42 \\
\hline
\hline
\end{tabular}
\end{center}
\end{table}

To trace the incremental localization of 5$f$ electrons from Am to Cf, we analyze the kinetic energy ($E_{\mathrm {kin}}$), potential energy ($E_{\mathrm {pot}}$), spin susceptibility ($\chi_{\mathrm {S0}}$) and charge susceptibility ($\chi_{\mathrm{D}}$) for dhcp and fcc phases, as listed in Table~\ref{tab:energy}. We begin with the evolution of energy scales. From Am to Cf, the absolute magnitude of the kinetic energy $E_{\mathrm {kin}}$ remains consistently small, on the order of 0.7 $\sim$ 0.8~eV for dhcp phase and 0.6 $\sim$ 1.2~eV for fcc phase. This reflects the intrinsically narrow bandwidth of the 5$f$ electrons, a prerequisite for strong correlation effects. As the 5$f$ electron count rises from Am to Cf, the kinetic energy shows a gradual but systematic decrease, signaling a progressive suppression of electron hopping, a direct indicator of enhanced 5$f$ electron localization. Meanwhile, for dhcp and fcc phases, the potential energy $E_{\mathrm {pot}}$ undergoes a dramatic increase, escalating from 67~eV in Am to 243~eV in Cf, representing a more than threefold rise. This surge stems from the growing Coulomb repulsion interaction and Hund's exchange interaction, both of which intensify as electrons become increasingly confined to the atomic site. As the 5$f$ shell fills, the concurrent decrease in kinetic energy and rapid rise in potential energy signify enhanced electron localization and strengthening electron correlations.

Magnetic response functions provide a complementary perspective on this localized state. The spin susceptibility $\chi_{\mathrm {S0}}$, which quantifies the system's tendency to form magnetic moments and respond to external fields, undergoes a qualitative transformation across the series. In Am, spin susceptibility $\chi_{\mathrm {S0}}$ is very small (0.296 for dhcp phase and 0.543 for fcc phase), consistent with a non-magnetic ground state dominated by strong spin-orbit coupling~\cite{PhysRevLett.94.097002,PhysRevB.76.073105}. A dramatic change occurs at Cm, where $\chi_{\mathrm {S0}}$ surges by more than two orders of magnitude, reaching 110.8 for the dhcp phase and 203.6 for the fcc phase. This abrupt surge signals the emergence of robust local moments, in agreement with magnetic susceptibility measurements on Cm and Bk metals, which report effective moments 7.58 $\mu_B$/atom for Cm-I and Cm-II~\cite{AmCmmag1975,PhysRevB.99.224419} and around 8 $\mu_B$/atom for Bk~\cite{RevModPhys.81.235}. In Cm, the 5$f^7$ configuration fills the $j = 5/2$ subshell and begins populating the $j = 7/2$ states. Hund's rule aligns the spins in parallel, yielding a half-filled shell with maximal spin. The resulting high spin susceptibility reflects the high responsiveness of these moments, which are easily polarized by external perturbations such as temperature or magnetic field. This trend intensifies in the heavier elements: $\chi_{\mathrm {S0}}$ climbs to 553.6 (dhcp) and 354.9 (fcc) in Bk, and further to 769.7 (dhcp) and 811.1 (fcc) in Cf. The continued growth of spin susceptibility confirms that local moments not only persist but become increasingly dominant as the 5$f$ count advances.

The charge susceptibility $\chi_{\mathrm{D}}$ offers insight into the evolution of orbital degrees of freedom. In Am, $\chi_{\mathrm{D}}$ is positive (0.01 for dhcp and fcc phases), indicating the absence of orbital order or magnetic frustration. Beginning with Cm, however, $\chi_{\mathrm{D}}$ for the dhcp phase abruptly turns negative, taking values of –3.16, –1.30, and –3.60 for Cm, Bk, and Cf, respectively. A similar trend is observed for the fcc phase, with values of –0.29, –4.92, and –4.42 across the same elements. This sign reversal is a key fingerprint of emerging orbital-selective correlations. A negative charge susceptibility reflects either a partial quenching of orbital angular momentum or the development of a preferential orbital occupation. In the present context, this corresponds to the progressively differentiated roles of the $j$ = 5/2 and $j$ = 7/2 subshells. The negative values observed from Cm onward directly confirm the onset of orbital selectivity, meaning that the $j$ = 5/2 and $j$ = 7/2 manifolds begin to participate differently in both local moment formation and the scattering of itinerant electrons. In summary, increasing 5$f$ electron count drives the system from Am, a non-magnetic metal, to Cm and Bk, where local moments and orbital selectivity govern the low-energy physics. The systematic decrease in kinetic energy, the explosive rise in potential energy and spin susceptibility, and the sign reversal of charge susceptibility all lead to a conclusion that higher 5$f$ occupancy gradually enhances electron localization, correlation strength, and orbital differentiation.

\section{Conclusions\label{sec:summary}}
In this study, the electronic structures of Am, Cm, Bk, and Cf in both dhcp and fcc phases have been comprehensively investigated using density functional theory combined with the embedded dynamical mean-field theory. From Am to Cf, the 5$f$ electrons exhibit progressive localization, accompanied by weakened valence state fluctuations, enhanced correlation strength, and the emergence of orbital-selective correlations between the $j$ = 5/2 and $j$ = 7/2 manifolds. Am exhibits moderate correlation strength and a nonmagnetic ground state. In contrast, Cm and Bk fall into a strongly correlated localized regime, characterized by Hubbard band formation and substantial effective electron masses. Strikingly, Cf displays the strongest orbital-selective correlations in the series. Although the fcc structure generally amplifies correlation effects relative to the dhcp phase, the bonding behavior of the 5$f$ electrons remains largely similar across both structures, underscoring a close link between 5$f$ electron correlation and lattice stability. The competition between exchange interaction and spin–orbit coupling, driven by variations in 5$f$ occupancy across the $j$ = 5/2 and $j$ = 7/2 manifolds, leads to distinct angular momentum coupling schemes. When exchange interaction prevails, as in Cm and Bk, the system adopts an intermediate coupling scheme biased toward LS coupling. Under Hund's rule, the 5$f$ spins align in parallel, resulting in magnetic ground states. When spin-orbit coupling dominates, as in Am and Cf, the system is governed by $jj$ coupling and manifests nonmagnetic ground states. Analyses of energy, spin and charge susceptibility further corroborate the progressive localization of 5$f$ electrons and the emergence of orbital-selective correlations from Am to Cf. Our findings therefore provide a unified description of 5$f$ electron evolution from Am to Cf, linking electronic structure, magnetic ground states, and lattice stability, while laying the groundwork for understanding heavy-actinide behavior under extreme conditions and guiding future actinide-based applications.

\begin{acknowledgments}
This work is supported by the National Natural Science Foundation of China (under Grant No.~12474241), the National Key Research and Development Program of China (under Grant No.~2024YFA1408600), and the Presidential Foundation of CAEP (under Grant No.~YZJJZQ2024014).
\end{acknowledgments}

\bibliography{dhcp}

\end{document}